\documentclass[aps,twocolumn, amsmath,amssymb,pra,superscriptaddress]{revtex4-2}

\usepackage{enumerate}
\usepackage{graphicx}
\usepackage{dcolumn}
\usepackage[dvipsnames]{xcolor}
\usepackage{bm}

\usepackage{empheq,etoolbox}
\usepackage{bbold}

\usepackage{color}
\usepackage[dvipsnames]{xcolor}
\usepackage{bm}
\usepackage{setspace}
\usepackage{adjustbox}
\usepackage[normalem]{ulem}
\usepackage{verbatim}
\usepackage{gensymb}
\usepackage{tabularx}
\usepackage{placeins}
\usepackage[mathlines]{lineno}
\usepackage{comment}
\usepackage[a4paper, total ={6in,8in}]{geometry}

\begin{document}

\title{Vortex Lines in Ultralight Bosonic Dark Matter around Rotating Supermassive Black Holes}





\author{K. Korshynska}
\affiliation{Department of Physics, Taras Shevchenko National University of Kyiv, 
64/13, Volodymyrska Street, Kyiv 01601, Ukraine}
\affiliation{Physikalisch-Technische Bundesanstalt (PTB), Bundesallee 100, D-38116 Braunschweig, Germany}
\affiliation{Institut für Mathematische Physik, Technische Universität Braunschweig, Mendelssohnstraße 3, 38106 Braunschweig, Germany}
\author{O.O. Prykhodko}
\affiliation{Department of Physics, Taras Shevchenko National University of Kyiv, 
64/13, Volodymyrska Street, Kyiv 01601, Ukraine}
\author{E.V. Gorbar}
\affiliation{Department of Physics, Taras Shevchenko National University of Kyiv, 
64/13, Volodymyrska Street, Kyiv 01601, Ukraine}
\affiliation{Bogolyubov Institute for Theoretical Physics, 14-b Metrolohichna Street, Kyiv 03143, Ukraine}
\author{Junji Jia}
\affiliation{School of Physics and Technology, Wuhan University, 
299 Bayi Road, Wuhan, 
Hubei Prov., China 430072}
\author{A.I. Yakimenko}
\affiliation{Department of Physics, Taras Shevchenko National University of Kyiv,
64/13, Volodymyrska Street, Kyiv 01601, Ukraine}
\affiliation{Dipartimento di Fisica e Astronomia ’Galileo Galilei’,
Universit{\'a} di Padova, via Marzolo 8, 35131 Padova, Italy}

\begin{abstract}
 Theoretical analysis of the interaction between superfluid dark matter and rotating supermassive black holes offers a promising framework for probing quantum effects in ultralight dark matter and its role in galactic structure. We study how black hole rotation influences the state of ultralight bosonic dark matter, focusing on the stability and dynamics of vortex lines. The gravitational effects of both dark matter and the black hole on the physical properties of these vortex lines, including their precession around the black hole, are analyzed.


Keywords: {dark matter, ultralight bosons, Bose-Einstein condensate, supermassive Kerr black hole, gravitoelectromagnetism}
\end{abstract}

\maketitle

\section{Introduction}
\label{sec: Introduction}

Nature of dark matter (DM)
remains a mystery of modern physics.
On large scales, i.e., for distances much larger than the typical galaxy size, astrophysical observations can be successfully explained by the cold dark matter model (CDM), which assumes DM to be a collisionless sufficiently cold, i.e., non-relativistic, perfect fluid \cite{Ferreira, jackson2023search}. However, the CDM encounters the cusp-core, missing satellites, and too-big-to-fail problems at smaller scales such as the size of a typical galaxy. 

Ultralight dark matter (ULDM) models assume that DM consists of ultralight bosons  \cite{Ferreira}. These models reproduce CDM phenomenology on large scales and 
naturally suppress small-scale structures. ULDM models are characterized by the presence of cores and dynamic effects that arise from the Bose-Einstein condensate (BEC) formed in the central regions of galaxies \cite{B_hmer_2007, https://doi.org/10.48550/arxiv.1611.09610,PhysRevD.86.064011, Hui_2017, rindler2014complex, sikivie2009bose}. ULDM models have been widely investigated via cosmological simulations \cite{Schive_2014, Matos_2000, PhysRevD.62.103517} and their theoretical predictions were compared with observational data from the rotation curves of galaxies \cite{bar2018galactic, de2020dynamical}, the stellar kinematics measurements in dwarf galaxies \cite{goldstein2022viability}, polarimetry with axions \cite{gan2024detecting}, observations of collisions of clusters of galaxies \cite{Harvey_2015, 2008arXiv0805.3827L} and constraints from black hole (BH) superradiance \cite{baryakhtar2021black, collaviti2024observational, armengaud2019supermassive}. 

The Gross-Pitaevskii equation for ULDM models with dissipation predicts that the ULDM forms a BEC core and an isothermal halo with an effective temperature \cite{chavanis2019predictive, chavanis2022heuristic,launhardt2002nuclear, schonrich2015kinematic, portail2016dynamical, Schive_2014-core-halo}. In this paper, we focus on the central BEC core, where a supermassive BH is located in typical massive galaxies \cite{abuter2023polarimetry, merritt2013dynamics}, which slightly perturbs the overall BEC core distribution except the most central region. Such perturbation within the framework of ULDM was discussed in \cite{chavanis2019mass}, however, only the Schwarzschild BH was investigated.

It is well known that the BEC manifests superfluidity. Although superfluid cannot rotate as a whole, it admits the formation of vortices with vanishing BEC wave function at the vortex line and quantized circular flow around the vortex line 
\cite{Ferreira}. In the case of galactic BEC, such vortex structures with nonzero angular momentum and phase dislocation at the vortex core may be self-sustained \cite{Chavanis, Hui_2021, Nikolaieva, Dmitriev} or induced by external rotation \cite{zhang2018slowly, rindler2012angular, madarassy2013numerical}. In \cite{silverman2002dark, zinner2011vortex, kain2010vortices, rotha2002vortices} it was shown that rotation of spiral galaxies would cause vortex lattices to form in ULDM haloes. In the central region of a galaxy with the supermassive Kerr BH, such rotation is also introduced by the BH. Therefore, the main question addressed in this study is how BEC vortices form in the presence of a rotating BH and what is their dynamics. 

To describe the gravitational field of a supermassive Kerr BH and DM in the BEC state, we employ the well-known approach of gravitoelectromagnetism \cite{Toth,Hehl,Medina,Mashhoon,Wald,sarkar2018gravitationally}, which is convenient to describe the rotating BH metric \cite{Wald} sufficiently far from the BH.
The Gross-Pitaevskii-Poisson system of equations with the self-induced and BH gravitoelectromagnetic potentials allows us to investigate the dynamics of BEC in BH background and account for the impact of BH rotation on vortices formed in DM. In particular, we will analyze how a rotating BH induces vortex lines which may appear in DM.

We note that the post-Newtonian corrections taken into account do not allow to explore all possible physical regimes of the combined BH and BEC system. A fully relativistic analysis of the Einstein-Klein-Gordon field equations performed in  \cite{cardoso2022parasitic, duque2023axion, mitra2023probing} has shown that the BH is expected to swallow the DM soliton, which would eventually cause dramatic changes in the system. In our study, we consider such time scales for which the process of the DM inspiral does not sufficiently perturb the state of BEC and BH. Estimates of these time scales are given in \cite{bar2018galactic}.


The model equations (\ref{modeleq1})-(\ref{modeleqlast}) are similar to the Ginzburg-Landau equations in the theory of superconductivity, where the Cooper pairs condensate corresponds to BEC and the magnetic field induced by magnetic dipole corresponds to the gravimagnetic field due to the rotating supermassive BH in the problem under consideration. For magnetic dipole placed inside superconductor, it was found that the Cooper pairs condensate can form complex structures leading to the formation of vortex lines \cite{doria2007effect}. This analogy between the magnetic dipole problem in the theory of superconductivity and
BEC in the presence of rotating supermassive BH suggests that similar complex structures of vortex lines could form in BEC in the vicinity of rotating supermassive BH and partially motivates the present study.

The paper is organized as follows. The model of ULDM in the Kerr BH background is formulated in Sec.\ref{sec:model}. Stationary states of the ground state soliton and vortex lines are studied in Sec.~\ref{sec: Stationary Vortex States}. The energy analysis and dynamics of vortex lines are considered in Sec.~\ref{sec:Energy analysis and dynamical stability}. Conclusions are drawn in Sec.~\ref{sec:conclusions}. The Kerr BH metric and Gross-Pitaevskii equation describing DM are discussed in Appendices~\ref{App: Notes on spacetime metric} and \ref{App: Derivation of GP equation}, respectively. In Appendix~\ref{Appendix: Model restrictions}, we discuss in more detail some assumptions made in the formulation of our model. Dynamics of a vortex line is described in Appendix~\ref{App: Dynamics of a vortex line}.

\section{Model}
\label{sec:model}

Before formulating equations which describe a supermassive BH and ULDM, it is worthwhile to specify the set of the most relevant parameters characterizing the considered system. These parameters are:
\begin{itemize}
    \item half of BH Schwarzschild radius $R_\textrm{S} = GM_\textrm{BH}/c^2$ defining mass and gravielectric potential of supermassive BH;
    \item parameter of BH rotation $a = J/(cM_\textrm{BH})$ specifying angular momentum and gravimagnetic field of BH;
    \item DM particle mass $m$;
    \item s-wave scattering length of a DM particle $a_\textrm{s}$, characterizing the weak self-interaction in the BEC core;
    \item mass of the BEC core $M_\textrm{DM}$.
\end{itemize}

In the following we will use coordinate system $\mathbf{r} = \{x,y,z\}$ with the center located at BH position and $z$-axis directed along BH angular momentum $\mathbf{J}$. To account for the symmetry, we will also use the corresponding cylindrical \mbox{$\mathbf{r} = \{r_\perp = \sqrt{x^2 + y^2}, \phi = \textrm{sgn}(y)\arccos(x/r_\perp), z\}$} and  spherical \\ \mbox{$\mathbf{r} = \{r = \sqrt{r_\perp^2 + z^2},\theta = \arccos(z/r), \phi \}$} \\ coordinates.

\subsection{Spacetime metric and geodesics}

To determine the gravitational field of DM core and a supermassive Kerr BH we employ the well-known  gravitoelectromagnetism (GEM) approach \cite{Wald} which was previously applied to galactic structures in \cite{Toth,Medina,Mashhoon}. This formalism is derived from Einstein’s field equations in the case of slowly moving matter and was used for the description of the vortex BEC core in \cite{korshynska2023dynamical}. For the reader`s convenience, the exact spacetime metric of Kerr BH is presented in Appendix~\ref{App: Notes on spacetime metric}. In the linear order of BH and DM contributions, we have the following spacetime metric:
\begin{multline}
  ds^{2} = g_{\mu \nu}dx^{\mu}dx^{\nu} \\= (\eta_{\mu \nu} + \gamma_{\mu \nu}^\textrm{DM} + \gamma_{\mu \nu}^\textrm{BH})dx^{\mu}dx^{\nu}  = \\
  = \left(1 - \frac{2R_\textrm{S}}{r} + \frac{2\Phi^\textrm{DM}}{c^{2}}\right)(dx^{0})^{2}  \\ -\frac{4R_\textrm{S} a \sin^{2}\theta}{r^{2}}dx^{0} rd\phi + \frac{4}{c^2}(\mathbf{A}^\textrm{DM}\mathbf{dx})dx^0  \\ - \left[1 + \frac{2R_\textrm{S}}{r} + \left(\frac{2R_\textrm{S}}{r}\right)^{2}  - \frac{a^{2}\sin^{2}\theta}{r^{2}} - \frac{2\Phi^\textrm{DM}}{c^{2}} \right]dr^{2} \\+ \left(-1 - \frac{a^{2}}{r^{2}}\cos^{2} \theta + \frac{2\Phi^\textrm{DM}}{c^{2}}\right)r^{2}d \theta^{2} \\ + \left(-1 - \frac{a^{2}}{r^{2}} + \frac{2\Phi^\textrm{DM}}{c^{2}}\right)r^{2} \sin^{2} \theta d \phi^{2}.
  \label{eq: spacetime metric}
\end{multline}

The acceleration of a classical probe particle is defined by the gravielectric potential $\Phi = \Phi^\textrm{DM} - c^{2}\frac{R_\textrm{S}}{r}$ and the gravimagnetic vector potential $\mathbf{A} = \mathbf{A}^\textrm{DM} + \mathbf{A}^\textrm{BH}$ where $\mathbf{A}^\textrm{BH} = - c^{2}\frac{aR_\textrm{S}}{r^{2}}\sin \theta \mathbf{e}_{\phi}$, and equals
\begin{equation}
    \mathbf{a} = -\nabla \Phi - \frac{2}{c}[\mathbf{v} \times [\nabla \times \mathbf{A}]].
    \label{eq: GEM classical}
\end{equation}
Clearly, to specify completely this acceleration, we should determine $\Phi^\textrm{DM}$ and $\mathbf{A}^\textrm{DM}$. For this, we should consider the equations which govern the evolution of ULDM.

\subsection{ULDM in the curved spacetime}
\label{subsec: ULDM in the curved spacetime}

In the non-relativistic limit, this evolution is defined by the following Gross-Pitaevskii (GP) equation:
\begin{eqnarray}
    i\hbar \frac{\partial \psi}{\partial t}  &=& -\frac{\hbar^{2}}{2m}\Delta \psi + m\tilde{\Phi} \psi - 2\frac{i\hbar}{r}\frac{A_{\phi}^\textrm{BH}}{c \sin \theta}\frac{\partial \psi}{\partial \phi}\nonumber \\
    &+& gN |\psi|^{2} \psi,\label{modeleq1}\\
    \mathbf{j} = &-&\frac{i\hbar}{2m}\left[\psi^{*}\nabla \psi - \psi \nabla \psi^{*}\right] + \frac{2}{c}\mathbf{A}^\textrm{BH}|\psi|^{2},\label{modeleq2}
\end{eqnarray}
where
$\tilde{\Phi} = \Phi - \frac{c^{2}}{2}\left(\frac{2R_\textrm{S}}{r}\right)^{2}$ includes the post-Newtonian correction of the second order. The key steps of derivation of the Eq.~(\ref{modeleq1}) are presented in Appendix~\ref{App: Derivation of GP equation}. Here and in what follows we will focus on the impact of BH gravimagnetic field $A^\textrm{BH}$ and neglect the effect of $\mathbf{A}^\textrm{DM}$ (validity of this assumption is discussed in Appendix~\ref{Appendix: Model restrictions}). To close the system of equations, we recapitulate that gravielectric and gravimagnetic potentials induced by DM are defined by the following Poisson and Ampere equations \cite{Wald}:
\begin{eqnarray}
    \Delta \Phi^\textrm{DM} &=&  4\pi G N m |\psi|^{2} \label{modeleqPoisson},\\
    \Delta \mathbf{A}^\textrm{DM} &=& \frac{8 \pi G N m}{c}\mathbf{j}
    \label{modeleqlast}.
\end{eqnarray}
Here $|\psi|^{2}$ denotes the particle concentration, so that $\rho = mN|\psi|^{2}$ is the BEC mass density. The coupling strength depends on the scattering length $a_\textrm{s}$ and the particle mass $m$ via $g = 4\pi a_\textrm{s} \hbar^2/m$. Note that equations (\ref{modeleq1}) and (\ref{modeleq2}) are similar to the Ginzburg-Landau equations in the theory of superconductivity. The correlation length reads $\xi = 1/\sqrt{8\pi a_\textrm{s}\rho/m}$, which defines the size of a vortex core in the BEC.

The gravimagnetic field in the GP equation (\ref{modeleq1}) leads to a term similar to that due to external rotation and equals
\begin{multline}
   - 2\frac{i\hbar}{r}\frac{A_{\phi}^\textrm{BH}}{c \sin \theta}\frac{\partial \psi}{\partial \phi} = - \mathbf{\Omega} \cdot [\mathbf{r} \times \hat{\mathbf{p}}] \psi \\= i
   \hbar\left(-\frac{1}{r}v_\phi \right) \left(-\frac{1}{\sin \theta}\frac{\partial}{\partial \phi}\right)\psi = \frac{i \hbar v_\phi}{r \sin \theta}\frac{\partial \psi}{\partial \phi}.
\end{multline}
We find that the corresponding velocity $v_\phi = -2A_\phi^\textrm{BH}(r, \theta)/c$ is coordinate dependent as well as the angular velocity $\mathbf{\Omega} =  caR_\textrm{S}/r^3 \mathbf{e}_z$. This velocity is non-relativistic $v_\phi \ll c$ in our setup $r/R_\textrm{S} \gg 1$.

\subsection{Dimensionless equations}
\label{sec:ansatz}

For further analysis, it is convenient to use the fact that the GPP system of equations is invariant under the transformation $t = \lambda_{*}^{2}t'$, $\mathbf{r} = \lambda_{*}\mathbf{r}'$, $\psi = \lambda_{*}^{-2}\psi'$, $\Phi = \lambda_{*}^{-2}\Phi'$, $g = \lambda_{*}^{2}g'$, where $\lambda_{*} > 0$, which allows us to rescale the coupling constant to $g=1$ \cite{Chavanis}. Then, in terms of dimensionless variables and wave function, we have the following equations:
\begin{align} 
&  i\frac{\partial\psi}{\partial t} = \left(-\frac{1}{2}\nabla^{2} + |\psi|^{2} + \Phi + i \Omega(r) \frac{\partial}{\partial \phi}\right)\psi,
  \label{GP equation dimensionless}\\
& \nabla^{2}\Phi^\textrm{DM} = |\psi|^{2}.
  \label{Poisson equation dimensionless}
\end{align}
 For current $\mathbf{j}$, the dimensionless version of Eq.~(\ref{modeleq2}) is
\begin{equation}
    \mathbf{j} =  -\frac{i}{2}(\psi^*\nabla \psi - \psi \nabla \psi^*) - r_\perp\Omega|\psi|^2 \mathbf{e}_\phi.
\end{equation}
The dimensionless variables are related to the dimensional ones as follows: $\lambda = 8\pi a_\textrm{s}/\lambda_\textrm{c}$, $\mathbf{r} = \mathbf{r}_{\mathrm{ph}}/{L}$, $t = \omega_{*} t_{\mathrm{ph}}$, $\Omega(r) = \Omega_\mathrm{ph}(r)/\omega_*$, $\Phi= \left(\frac{L}{\lambda_{\mathrm{c}}}\right)^{2}\frac{\Phi_{\mathrm{ph}}}{c^{2}}$, and $\psi = \frac{\lambda}{8\pi}\left(\frac{m_{\mathrm{Pl}}}{m}\right)^{2}\sqrt{4\pi GM}\frac{\hbar}{mc^{2}}\psi_{\mathrm{ph}}$. Here and in the following $\lambda_\textrm{c} = \hbar/(mc)$ stands for the Compton wavelength of a DM boson. Further, the distance and time scaling parameters are $L = \lambda_{\mathrm{c}}\frac{m_{\mathrm{Pl}}}{m}\sqrt{\frac{\lambda}{8\pi}} = \frac{m_{\mathrm{Pl}}\hbar}{m^{2}c}\sqrt{\frac{\lambda}{8\pi}}$ and $\omega_{*} = \frac{c\lambda_{\mathrm{c}}}{L^{2}}$. Rotation $\Omega(r)$ is introduced by BH gravimagnetic field, is coordinate dependent, and reads $\Omega(r) = \frac{2aR_\textrm{S}}{\lambda_\textrm{c} L}1/r^3$. The unit of energy reads $\epsilon = (\hbar^2/4\pi m_{Pl}\lambda_\mathrm{c}^2)(8\pi/\lambda)^{3/2}$.

Dimensionless gravielectric potential of the BEC and BH system reads
\begin{equation*}
\Phi(\mathbf{r}) =\Phi^\textrm{DM}(\mathbf{r}) - \frac{LR_\textrm{S}}{\lambda_\textrm{c}^2}\frac{1}{r} - 2\left(\frac{R_\textrm{S}}{\lambda_\textrm{c}}\right)^2\frac{1}{r^2},
\end{equation*}
where the last two terms describe dimensionless gravielectric potential induced by the BH with $r$ measured in units of $L$.

While the wave function in dimensional units was normalized to the total number of particles $N$, the normalization of dimensionless wave function $N_{0}$ is defined by the core mass
\begin{equation}
    \int |\psi|^2 d\mathbf{r} =  N_{0} = 4\pi\frac{M_\textrm{DM}}{m_\mathrm{Pl}}\sqrt{\frac{\lambda}{8\pi}},
    \label{eq: normalization}
\end{equation}
where $m_\textrm{Pl} = \sqrt{\frac{\hbar c}{G}}$ is the Planck mass and $\lambda/(8\pi)$ is the self-interaction coupling constant.


\subsection{Parameters and their numerical values}

\label{subsec: Parameter space}

To proceed, we should specify first numerical values of our parameters. In our model, we consider only distances $r \gg R_\textrm{S}$, therefore, $R_\textrm{S}/L \ll 1$. This also automatically assures the smallness of the post-Newtonian corrections. 

These parameters also define the dimensional model. Still it remains to choose the value of boson mass $m$. If it is fixed, then in view
\begin{equation}
 \frac{\lambda_\textrm{c}}{L} = \frac{m}{m_\textrm{Pl}}\sqrt{\frac{8\pi}{\lambda}}
\end{equation}
the self-interaction coupling constant $\lambda$ will be defined too. Together with $R_\textrm{S}/L$ and $N_0$ it determines the three observable physical parameters of the model: BEC core mass $M_\textrm{DM}$, BEC core radius in the absence of BH $R$, and BH mass $M_\textrm{BH}$
\begin{eqnarray*}
    M_{\textrm{DM}} &=& \frac{m_\textrm{Pl}^2}{m} \frac{N_0}{4\pi} \frac{\lambda_\textrm{c}}{L},\\
    R &=& \pi \left(\frac{\lambda_\textrm{c}}{L}\right)^{-1} \frac{\hbar}{mc},\\
    M_\textrm{BH} &=& \frac{R_\textrm{S}}{L}\left(\frac{\lambda_\textrm{c}}{L}\right)^{-1}\frac{m_\textrm{Pl}^2}{m}.
\end{eqnarray*}

In our subsequent analysis, we fix $N_0 = 10^3$. Then the above relations can be simplified
 \begin{eqnarray}
    \frac{ M_{\textrm{DM}}}{M_\odot} &=& 1.06 \times 10^{-8} \frac{\lambda_\textrm{c}}{L} \frac{1 \textit{eV}}{m} \label{eq: DM mass},\\
    \frac{R}{1 \textit{pc}} &=& 2.01 \times 10^{-23} \left(\frac{\lambda_\textrm{c}}{L}\right)^{-1} \frac{1 \textit{eV}}{m} \label{eq: DM radius},\\
    \frac{M_\textrm{BH}}{M_\odot} &=& 1.3 \times 10^{-10} \frac{R_\textrm{S}}{L}\left(\frac{\lambda_\textrm{c}}{L}\right)^{-1}\frac{1 \textit{eV}}{m}. \label{eq: BH mass}
\end{eqnarray}
The relations explicitly show that the impact of BH is more significant for large values of $R_\textrm{S}/L$ and small values of $\lambda_\textrm{c}/L$.

\begingroup
\setlength{\tabcolsep}{10pt} 
\renewcommand{\arraystretch}{1.2} 
\begin{table}[h]
\begin{tabular}{ |c|c|c| } 
 \hline 
 Parameter & Case I & Case II \\  
 \hline
 $M_\textrm{DM}/M_\odot$ & $1.06 \times 10^{10}$ & $1.06 \times 10^{12}$ \\ 
 $R/$pc & $2.01 \times 10^1$ & $2.01 \times 10^{-1}$ \\ 
 $M_\textrm{BH}/M_\odot$ & $1.3 \times 10^8$ & $1.3 \times 10^{11}$ \\
 $L$/pc & 6.4 & $6.4 \times 10^{-2}$ \\
 $\omega_*^{-1}$/yr. & $2.1 \times 10^4$ & 2.1\\
 $\lambda_\textrm{c}/L$  & $10^{-3}$ & $10^{-1}$\\
 $R_\textrm{S}/L$  & $10^{-6}$ & $10^{-1}$ \\
 \hline
\end{tabular}
\caption{Two parameter sets of DM galactic core and supermassive BH. Case I - DM with BH gravitational field in the Newtonian approximation, Case II - DM with BH gravitational field which includes the post-Newtonian correction }
\end{table}
\endgroup

In our numerical calculations, we will assume boson mass $m = 10^{-21}$ eV and use the two sets of parameters (see Eqs.~(\ref{eq: DM mass})-(\ref{eq: BH mass})): Case I - DM with BH gravitational field in the Newtonian approximation, Case II - DM with BH gravitational field which includes the post-Newtonian correction (see Table~I).

We note that the non-relativistic model formulated by GPP equations (\ref{GP equation dimensionless}-\ref{Poisson equation dimensionless}) is valid only in case when the relativistic process of DM inspiral does not cause any dramatic change to the BEC and BH system on the characteristic time scale of GPP dynamics. This indeed holds true in the cases I and II described above (see Appendix~\ref{Appendix: Model restrictions}).

\section{Stationary Vortex States}
\label{sec: Stationary Vortex States}
\label{sec:Vortex density and velocity}

In what follows, we focus on stationary solutions $\psi(\mathbf{r}, t) = \psi(\mathbf{r}) e^{-i \mu t}$ of the system of Eqs.~(\ref{GP equation dimensionless}) and (\ref{Poisson equation dimensionless}), where $\mu$ is the dimensionless chemical potential of BEC.

\subsection{Ground state}
\label{subsec: ground state}

\begin{figure}
\centering
\includegraphics[width=.5\textwidth]{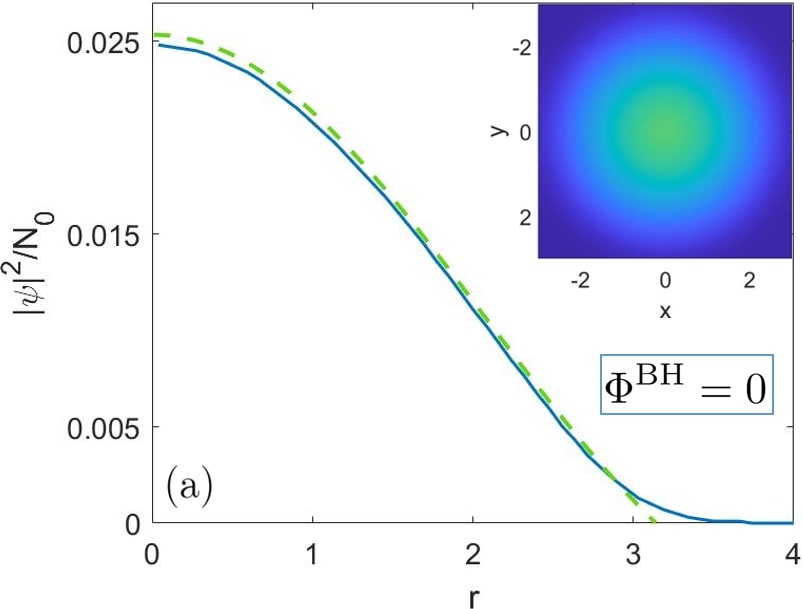}
\includegraphics[width=.5\textwidth]{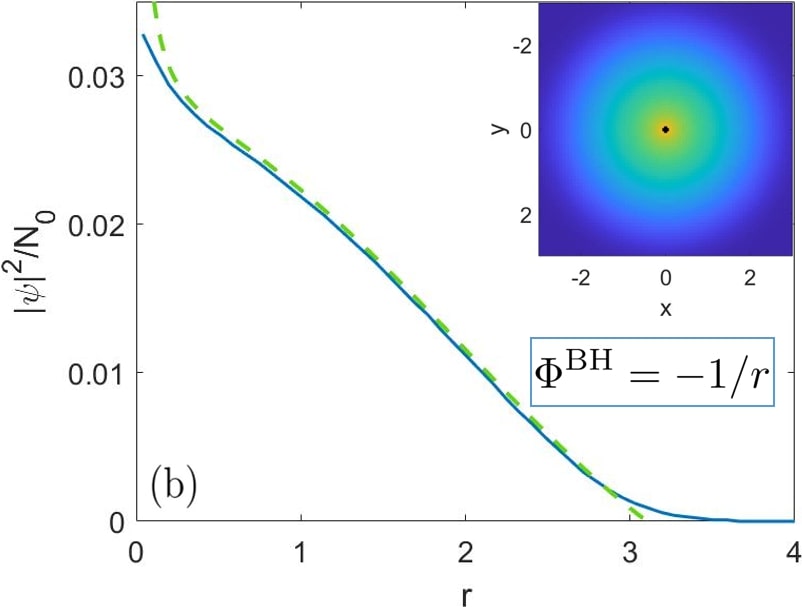}
\includegraphics[width=.5\textwidth]{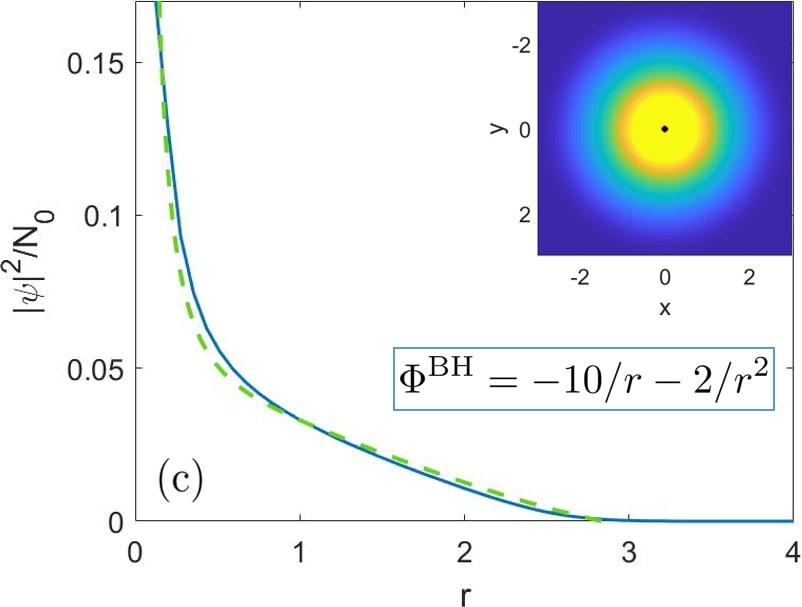}
   \caption{Comparison of numerical (solid blue line) and analytical Thomas-Fermi (green dashed line) solutions $|\psi(r)|^2/N_0$ to the GPP equations. Plots (a)-(c) represent the cases with no BH gravielectric potential, the Newtonian BH gravielectric potential, and the BH gravielectric potential with the first post-Newtonian correction. The insets represent gradient plots of $|\psi(x, y, z=0)|^2/N_0$ with the BH location depicted by a black dot.}
   \label{Fig: s=0 TF vs Num all}
\end{figure}

At distances far from the vortex axis the BEC density tends to that of the ground state solution. Let us briefly recapitulate the case of DM without BH. In the absence of BH, we can apply the Thomas-Fermi approximation (neglecting the kinetic energy term in the dimensional GPP equations (\ref{modeleq1}) and (\ref{modeleq2})) for the dimensionless BEC core density \cite{Chavanis}
\begin{equation}
 |\psi_0(r)|^2 =   
\frac{N_0}{4\pi^2} \begin{cases}
 \frac{\sin (\pi r/R)}{\pi r/R}, \,\, \,\,  r \leq R\\    0, \,\,\,\,  r > R ,
\end{cases}
\label{eq: TF density}
\end{equation}
where $R = R_\textrm{ph}/l= \pi(a_\textrm{s}\hbar^2/(Gm^3))^{1/2}/L  = \pi$. For example, in the Milky Way galaxy, this radius is $R = 1$ kpc \cite{chavanis2019predictive, korshynska2023dynamical}.  Then, using the Poisson equation (\ref{modeleqPoisson}), we find the corresponding gravitational potential
\begin{equation*}
     \Phi^\textrm{DM} = -\frac{N_0}{4 \pi^2}\begin{cases}
 \left[1 + \frac{\sin(\pi r/R)}{\pi r/R}\right], \,\, \,\,  r \leq R\\    \frac{R}{r}, \,\,\,\,  r > R .
\end{cases}
\end{equation*}
The chemical potential of such a BEC in the TF limit reads $\mu = -N_0/(4\pi^2)$ and the corresponding BEC density is depicted in Fig.~\ref{Fig: s=0 TF vs Num all}(a).

Let us generalize this result to the case of BEC core with BH. The BEC ground state is fully determined by the trapping potential geometry in the Thomas-Fermi approximation \cite{Chavanis}. Neglecting all derivatives in Eq.~(\ref{GP equation dimensionless}), we obtain the following system of equations:
\begin{eqnarray}
    \Phi^\textrm{DM} + \Phi^\textrm{BH} + \rho &=& -|\mu|\\
    \Delta \Phi^\textrm{DM} &=& \rho,
\end{eqnarray}
where we denoted the BEC density $|\psi_0|^2 = \rho$ and the BH gravielectric potential $\Phi^\textrm{BH}(r) = - \frac{LR_\textrm{S}}{\lambda_\textrm{c}^2}1/r - 2\left(\frac{R_\textrm{S}}{\lambda_\textrm{c}}\right)^2 \times 1/r^2$. To simplify the equations above, we apply Laplacian operator $\Delta$ to the first equation and then find
\begin{equation}
    \rho + \Delta \Phi^\textrm{BH} + \Delta \rho = 0.
    \label{Eq: full TF with BH}
\end{equation}
Taking into account that $\Delta (1/r) = 4\pi \delta(r)$ and $\Delta (1/r^2) = 2/r^4$, for $r \neq 0$, we obtain the equation
\begin{equation*}
    \rho + \Delta \rho - 4\left(\frac{R_\textrm{S}}{\lambda_\textrm{c}}\right)^2 \frac{1}{r^4} = 0, \label{Eq: TF with BH}
\end{equation*}
whose general solution is
\begin{multline}
    \rho(r) = A\frac{\sin r}{r} + B \frac{\cos r}{r} \\ + 2\left(\frac{R_\textrm{S}}{\lambda_\textrm{c}}\right)^2\frac{1}{r^2}\left(1 + r \textit{Im}[e^{-ir}Ei(ir)]\right),
    \label{eq: ground state rho}
\end{multline}
where $Ei(ir) = i\pi/2 - \int_r^{+ \infty} \frac{\exp [it]}{t} dt$ \cite{Mathematica}. Following Ref.~\cite{chavanis2019mass}, we fix constant $B$ by taking the limit $r \rightarrow 0$ in Eq.~(\ref{Eq: full TF with BH}). We see that the density profile equals in this limit
\begin{equation*}
    \rho(r) \approx \frac{B}{r} + 2\left(\frac{R_\textrm{S}}{\lambda_\textrm{c}}\right)^2\frac{1}{r^2}.
\end{equation*}
Substituting it in Eq.~(\ref{Eq: full TF with BH}) gives $B = L R_\textrm{S}/\lambda_\textrm{c}^2$.

There remains only one unknown constant $A$ in Eq.~(\ref{eq: ground state rho}). We can define it imposing the cut-off radius $\tilde{R}$, where density vanishes, i.e.,
\begin{multline*}
    A = - \frac{LR_\textrm{S}}{\lambda_\textrm{c}^2}\cot \tilde{R} \\- 2\left(\frac{R_\textrm{S}}{L}\right)^2\frac{1 + \tilde{R}\textit{Im}[e^{-i\tilde{R}}Ei(i\tilde{R})]}{\tilde{R}\sin \tilde{R}}.
\end{multline*}
Together with the normalization condition it determines the BEC mass radius relation $N_0(\tilde{R})$ and, therefore, finally fixes all parameters in the density distribution. If the post-Newtonian $1/r^2$ contribution to the BH gravielectric potential is neglected, we have $A = -B/\tan \tilde{R}$ and \cite{chavanis2019mass}
\begin{equation}
N_0 = 4\pi \frac{LR_\textrm{S}}{\lambda_\textrm{c}^2} \left[\frac{\tilde{R}}{\sin \tilde{R}} - 1\right].
\label{eq: N0 R relation BH analytical}
\end{equation}
For the relevant case of weak BH gravitational field the TF radius is only slightly decreased by the BH gravitational field and we have
\begin{equation}
\tilde{R} \approx \pi - \frac{4\pi^2}{N_0} \frac{LR_\textrm{S}}{\lambda_\textrm{c}^2}.
\label{eq: TF R for weak Bh}
\end{equation}

In the general case, the relation between $\tilde{R}$ and $N_0$ can be found only numerically.

Note that all our considerations above are independent of the particular choice of parameters. To illustrate these results, we consider the two sets of parameters I and II defined in Table~I. In case I, the post-Newtonian contribution $\propto 1/r^2$ is negligible and we find the DM radius $\tilde{R} = 0.9877 \pi$ (see, Eq.~(\ref{eq: N0 R relation BH analytical})), which is shown in Fig.~\ref{Fig: s=0 TF vs Num all}(b). Alternatively, case II of post-Newtonian BH potential is represented in Fig.~\ref{Fig: s=0 TF vs Num all}(c). The exact TF radius of such a DM halo equals $\tilde{R} = 0.8452 \pi$.

\subsection{State with the vortex}
\label{subsec:Wave function near vortex core}

In the vicinity of the vortex line $|\mathbf{r} - \mathbf{r}_0| \lesssim \xi = \pi \sqrt{2/N_0} = 0.14$, it is convenient to shift the origin of the coordinate system to the vortex center at $\mathbf{r}_0$ and perform rotation to  new coordinates $\mathbf{r}'$ so that the new $z'$ axis coincides with the vortex line. Close to the vortex line the BEC density changes notably with distance. Therefore, the dominating contribution to the GP equation (\ref{GP equation dimensionless}) comes from spatial derivatives connected with the kinetic energy term. For a single-charged vortex $n = 1$, we recapitulate the well-known result $\psi = r_\perp'e^{i \phi'}$ in cylindrical coordinates $[r_\perp', \phi', z']$ \cite{jackson1999vortex}. The density profile of a vortex line can be well approximated with an intuitive analytical ansatz
\begin{equation}
\psi_1 = \alpha(N_0)\frac{r_\perp'e^{i\phi'}}{\sqrt{r_\perp'^2 + (2\xi)^2}}\psi_0(r),
\label{eq: vortex line ansatz}
\end{equation}
which at small distances $r_\perp' \lesssim \xi$ reproduces behavior of the near-vortex-core solution, while at larger distances recovering the BEC ground state wave function $\psi_0$. Here $\alpha(N_0)$ stands for normalization constant, which assures normalization (\ref{eq: normalization}). Explicit expressions for $\psi_0(r) = \psi_0(\sqrt{r_\perp^2 + z^2})$ in the post-Newtonian gravielectric field of BH are given in the previous subsection. While the density profile of the vortex is given by $|\psi_1|^2$, its velocity reads $\mathbf{v} = c \lambda_\textrm{c}/r_\perp' \mathbf{e}_{\phi'}$. Note that in spite of increasing velocity when approaching vortex core, the non-relativistic model can still be used because of the vanishing BEC density within the coherence length $\xi$.

For the case of a vortex line parallel to 
the $z$ axis and displaced from the axis by $\Delta$, the approximate wave function is determined by Eq.~(\ref{eq: vortex line ansatz}) with $r_\perp'$ and $\phi'$, which are polar coordinates in the vortex plane and equal
\begin{eqnarray}
    x'&=& x - \Delta\, ,\label{eq: first dashed}\\
    r_\perp'&=& \sqrt{y^2 + (x')^2}\, ,\\
    \phi'&=& \arctan \left[\frac{y}{x'}\right] \label{eq: last dashed}.
\end{eqnarray}
Clearly, $x'$ stands for the $x$-coordinate shifted with respect to the vortex axis. For an on-axis vortex line ($\Delta = 0$) in the discussed case of large $N_0$ (or, equivalently, small $\xi$), the normalization constant for $\psi_1$ equals approximately
\begin{equation}
    \alpha(N_0) = \left[1 + \frac{3}{2}\xi^2\left(1 - \frac{\pi}{2} \log 
    \left(\frac{1}{\xi}\right)\right)\right]^{-1/2},
    \label{eq: norma}
\end{equation}
where $\xi = \pi \sqrt{2/N_0}$.

\section{Energy analysis and dynamics of vortex lines}
\label{sec:Energy analysis and dynamical stability}

\subsection{Energy analysis}
\label{subsec: Energy analysis}

\begin{figure*}[htp!]
\includegraphics[width=\textwidth]
    {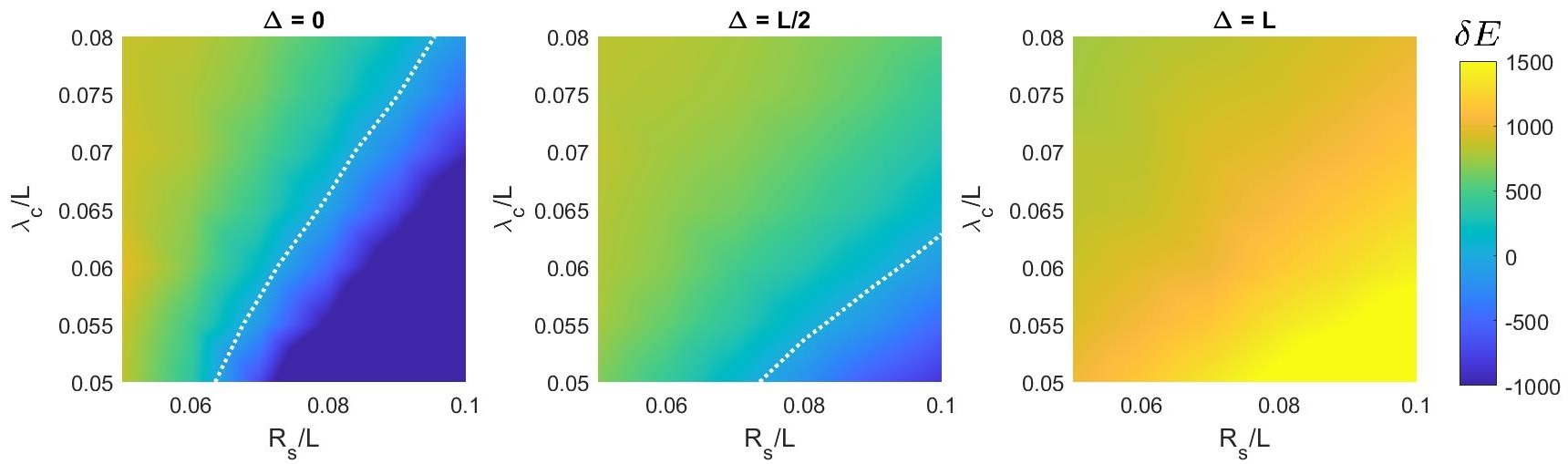}
   \caption{Energy difference $\delta E = E_\textrm{v} - E_0$ between the energy of a vortex line $E_\textrm{v}$, parallel to $z$ axis, and the energy $E_0$ of a ground state in the 2D parameter space $\{\lambda_\textrm{c}/L, R_\textrm{S}/L\}$. The three subplots differ by the distance $\Delta$ between the axis of  vortex line and $z$-axis. The white dotted contour corresponds to the line of $E_\textrm{v} = E_0$. The inner cut-off for the energy integration is $r = R_\textrm{S} = GM_\textrm{BH}/c^2$, which is the event horizon of an extreme Kerr BH.}
   \label{Fig: Energy analysis vortex line}
\end{figure*}

In this Section, we perform energy analysis of a vortex configuration in the 2D parameter space $\{R_\textrm{S}/L, \lambda_\textrm{c}/L\}$. For this we calculate the energy functional
\begin{multline}
    E [\psi] = \frac{1}{2} \int d \mathbf{r} |\nabla \psi|^2 + \frac{1}{2}\int d\mathbf{r}|\psi|^4  \\ + \int d\mathbf{r} \left(\frac{1}{2}\Phi^\textrm{DM} + \Phi^\textrm{BH}(r)\right) |\psi|^2 \\ + \frac{i}{2}\int d\mathbf{r} \Omega (r) \left(\frac{\partial \psi}{\partial \phi}\psi^* - \frac{\partial \psi^*}{\partial \phi}\psi\right),
    \label{eq: energy}
\end{multline}
corresponding to Eq.~(\ref{GP equation dimensionless}). Here $\Omega(r) = \frac{2aR_\textrm{S}}{\lambda_\textrm{c} L}1/r^3$. 

The vortex line (\ref{eq: vortex line ansatz}) is energetically favorable if its energy $E_\textrm{v}= E[\psi_1]$ is less than the ground state energy $E_0= E[\psi_0]$. The result of the energy calculation for the energy difference $\delta E = E_\textrm{v} - E_0$ in the case of extreme Kerr BH $a = R_\textrm{S}$ is given in Fig.~\ref{Fig: Energy analysis vortex line}. We see that the closer is the vortex line to the BH, the more energetically favorable it is. Therefore, the possibility of the formation of an on-axis vortex line ($\Delta = 0$) imposes the weakest condition on parameters $\lambda_\textrm{c}/L$ and $R_\textrm{S}/L$, namely, $\lambda_\textrm{c}/L < 0.9 R_\textrm{S}/L - 5.5 \times 10^{-3}$. On the contrary, the off-axis vortex line is less affected by the coordinate-dependent angular velocity $\Omega \propto 1/r^3$ of the BH and thus is less probable to be excited. For large displacement $\Delta$ (like $\Delta = L$ in Fig.~\ref{Fig: Energy analysis vortex line}) BH rotation no longer allows for the emergence of energetically favorable vortex lines.

\subsection{Dynamics of vortex lines}

\begin{figure*}[htp!]
\includegraphics[width=\textwidth]{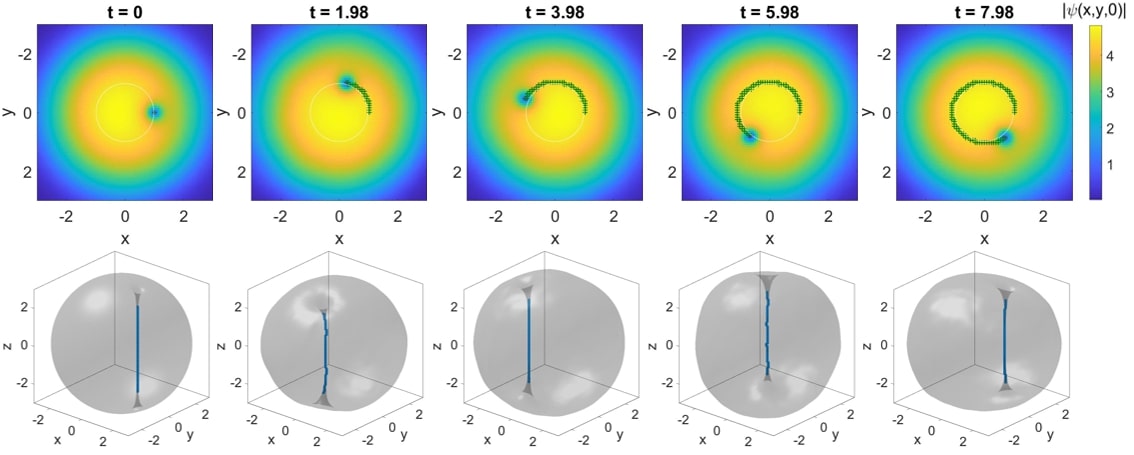}
   \caption{The dynamics of a vortex line displaced from the $z$-axis by $\Delta/L = 1$ in the form of snapshots at different moments of time $t$ in the absence of BH. Top row: gradient plot showing the absolute value of the wave function $|\psi(x,y,z=0)|$ in $z=0$ plane, white circle shows the isoline of ground state density at initial displacement, green crosses depict the trajectory of vortex line. Bottom row: the grey isosurface of $|\psi(x,y,z)|$ with the blue contour showing the vortex line.}
\label{Fig: dyn no BH}
\end{figure*}

\begin{figure*}[htp!]
\includegraphics[width=\textwidth]
    {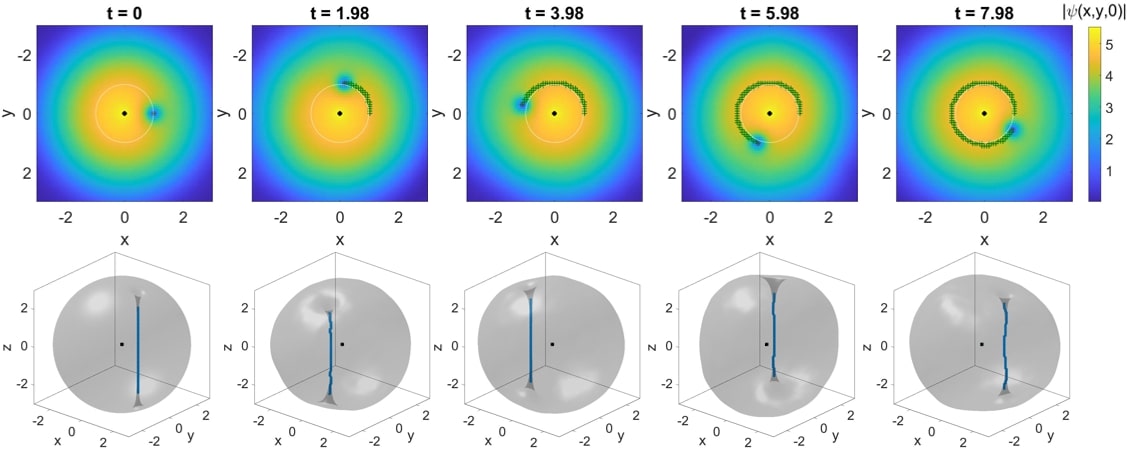}
   \caption{The dynamics of a vortex line displaced from the $z$-axis by $\Delta/L = 1$ in the form of snapshots at different moments of time $t$.  BH parameters are $R_\textrm{S}/L = 10^{-6}$ and $\lambda_\textrm{c}/L = 10^{-3}$ and its position is depicted by a black point at $\mathbf{r} = 0$. Top row: gradient plot showing the absolute value of the wave function $|\psi(x,y,z=0)|$ in $z=0$ plane, white circle shows the isoline of ground state density at initial displacement, green crosses depict the trajectory of vortex line. Bottom row: the grey isosurface of $|\psi(x,y,z)|$ with the blue contour showing the vortex line.}
   \label{Fig: dyn Newt BH}
\end{figure*}

\begin{figure*}[htp!]
\includegraphics[width=\textwidth]
    {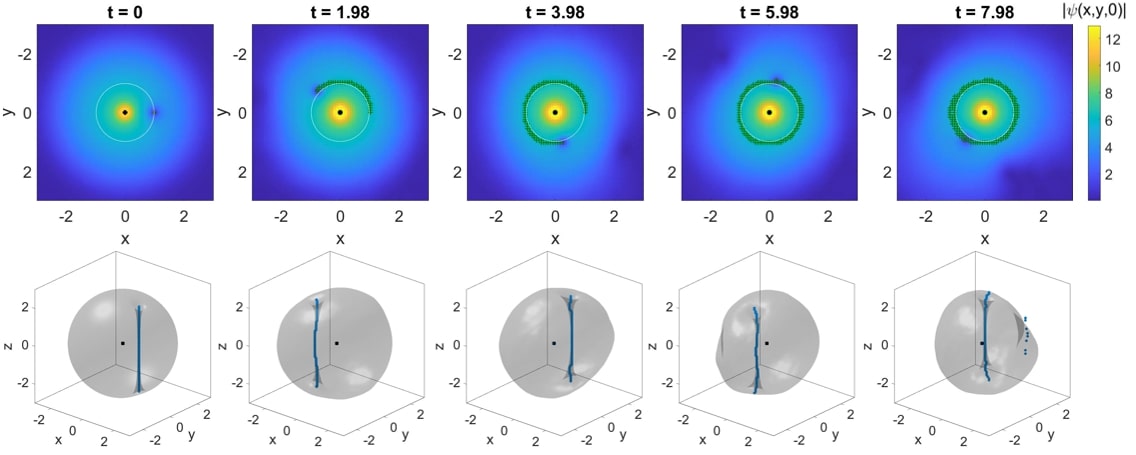}
   \caption{The dynamics of a vortex line displaced from the $z$-axis by $\Delta/L = 1$ in the form of snapshots at different moments of time $t$.  BH parameters are $R_\textrm{S}/L = 0.1$ and $\lambda_\textrm{c}/L = 0.1$ and its position is depicted by a black point at $\mathbf{r} = 0$. Top row: gradient plot showing the absolute value of the wave function $|\psi(x,y,z=0)|$ in $z=0$ plane, white circle shows the isoline of ground state density at initial displacement, green crosses depict the trajectory of vortex line. Bottom row: the grey isosurface of $|\psi(x,y,z)|$ with the blue contour showing the vortex line.}
   \label{Fig: dyn pNewt BH}
\end{figure*}

Let us analyze dynamics of a vortex line in ULDM. Depending on the initial displacement $\Delta$, a vortex line may exhibit different dynamics (see Figs.~3-5). An on-axis vortex line ($\Delta = 0$) remains stationary in time, as we show numerically in the 3D numerical simulations, similar to the studies in atomic BECs \cite{jackson1999vortex, mcgee2001rotational}. In contrast, an off-axis vortex line ($\Delta \neq 0$) is not stationary: it bends and follows a circular trajectory around the $z$-axis (see, Figs.~3-5). For a particular case of vortex line localized at $\Delta = L$,  we observe precession around the $z$-axis. It implies rotation along a path of constant energy, similar to the same effect studied in atomic BECs \cite{jackson1999vortex, mcgee2001rotational}. 

This precession is independent of rotation with angular velocity $\Omega (r)$ and appears due to the non-uniform BEC ground-state density (see, Eq.~(\ref{eq: ground state rho})). For an analytical estimate, we make use of the ansatz wave function (\ref{eq: vortex line ansatz}) and make a simplifying assumption that a straight vortex line remains straight with time, i.e., all points along the vortex axis move with the same velocity $\mathbf{v}(x,y)$. Then our original 3D problem is simplified to a problem of motion in the 2D $x-y$ plane. This assumption holds only approximately because numerical simulations show that the vortex line bends with time (see, 3D plots in Figs.~3-5). Still it allows us to understand the precession of a vortex line as a whole (see, 2D plots in Figs.~3-5). In addition, such an estimate does not take into account the occurrence of low-energy vortices on the BEC core outskirts (see, panel with $t = 7.98$ in Fig.~5), which leads to more complex regimes of dynamics discussed in \cite{asakawa2024corotation}.

\subsubsection{Self-gravitating DM in the absence of BH}

Numerical simulations for the DM halo with dimensionless mass $N_0 = 10^3$ reveal precession with period $T = 8.5$ at $\Delta = 2$, $T = 9.3$ at $\Delta = 1$, and $T = 9.5$ at $\Delta = 1/2$ (the case $\Delta = 1$ is shown in Fig.~3). For the parameter set I in the absence of BH, they give $T = 1.8 \times 10^5$ yr, $T = 1.9 \times 10^5$ yr, and $T = 2.0 \times 10^5$ yr for displacements $\Delta = 12.8$ pc, $\Delta = 6.4$ pc, and $\Delta = 3.2$ pc in dimensional units, respectively. We see that in agreement with similar studies in atomic BECs \cite{jackson1999vortex} we observe that the precession frequency increases with $\Delta$, so that vortex line precesses faster on the BEC outskirts. 

Now we will derive a rough and simple expression for the precession frequency following Refs.~\cite{jackson1999vortex, fetter2001vortices}, the key steps of the derivation are given in Appendix~\ref{App: Dynamics of a vortex line}. There we have shown that the vortex line follows the circular trajectory around the $z-$axis with radius $\Delta$ (equal to the initial displacement of the vortex line) and rotation frequency
\begin{equation}
    \omega_0(\Delta) = - \left[\frac{1}{4 \pi r_0 |\psi_0(r_0)|^2}\frac{\partial}{\partial r_0}\left(\int d^2\mathbf{r} |\psi|^4\right)\right]_{r_0 = \Delta},
    \label{eq: omega0}
\end{equation}
where $\psi_0(r) = \sqrt{\rho(r)}$ is the ground state wave function given in Eq.~(\ref{eq: ground state rho}). As to $\psi(\mathbf{r})$, we substitute the vortex wave function ansatz given by Eq.~(\ref{eq: vortex line ansatz}) and consider the case of large $N_0$ and small $\xi$.

For simplicity, we will find the analytical expression for $\omega_0$ in the regime $\Delta \ll 1$. Notice that in such a case the normalization constant $\alpha(N_0)$ does not change significantly with $r_0$, so we just take its constant value at zero displacement (\ref{eq: norma}), which in the lowest order of $\xi$, is equivalent to taking $\alpha(N_0) = 1$. 

For the case of self-gravitating DM in the absence of BH, $\psi_0(r)$ is given by Eq.~(\ref{eq: TF density}), which determines also $\psi(\mathbf{r})$ (\ref{eq: vortex line ansatz}) and after substitution in (\ref{eq: omega0}) leads to
\begin{equation}
    \omega_0 \approx  - \frac{3}{2} + \frac{4}{3}\log 
    \left(\frac{1}{\xi}\right)
    \label{eq: no BH omega 0}
\end{equation}
in the leading order of $\xi$, where $\xi = \pi \sqrt{2/N_0}$ is the dimensionless coherence length. Thus, we see that $\omega_0$ increases with $N_0$, this behavior is also evident from numerical simulations. For instance, in the discussed case $N_0 = 10^3$, Eq.~(\ref{eq: no BH omega 0}) gives the rotation period $T = 2\pi/\omega_0 = 5.63$, which is only a rough estimate of the corresponding numerical result.

\subsubsection{Self-gravitating DM in the presence of BH}

Let us now focus on the case of the Newtonian BH potential, which holds for the parameter set~I in Table~I. Numerical simulations in this case show that the vortex line displaced at $\Delta = 1$ and $\Delta = 2$  precesses with periods $T = 8.6$ and $T = 8.2$, respectively (see, Fig.~4). In physical units, for the parameter set I, these displacements are $\Delta = 6.4$~pc and $\Delta = 12.8$~pc, while the periods read $T = 1.8 \times 10^{5}$~yr and $T = 1.7 \times 10^{5}$~yr, respectively.

We can also obtain a rough analytical approximation of rotation frequency $\omega$ in the regime of self-gravitating BEC which is slightly perturbed by BH (this holds for the parameter set~I in Table~I). Mathematically, this can be formulated as accounting only for the Newtonian correction in Eq.~(\ref{eq: ground state rho}) and, moreover, assuming that $B =  L R_\textrm{S}/\lambda_\textrm{c}^2 \ll A$. In such a case, the TF radius of the BEC is defined by Eq.~(\ref{eq: TF R for weak Bh}) and, therefore, we obtain $A \approx N_0/(4\pi^2)$. Then, up to linear terms in $B/A$, we get
\begin{equation*}
    \omega = \frac{1}{1 + B/A \times 1/\Delta}\left(\omega_0 + \frac{\pi B}{16 \xi}\right),
\end{equation*}
where $\omega_0$ is given by Eq.~(\ref{eq: no BH omega 0}). For instance, for the parameter set I and displacement $\Delta = 0.1$, we obtain $2\pi/\omega = 3.49 < 2\pi/\omega_0$, while, for $\Delta = 0.02$, we have $2\pi/\omega = 7.43 > 2\pi/\omega_0$. This behavior illustrates the fact that the vortex line precesses faster in the lower density region of the BEC: while the presence of BH leads to the higher density peak in the central region, it leads to lower density on the BEC outskirts.

In the case where either the BH gravitational field is strong ($B/A \propto 1$) or the post-Newtonian correction is not negligible, there is no simple way to find approximate analytical expressions for $A$, $B$, and $\tilde{R}$ introduced in Sec.~\ref{subsec: ground state}. Therefore, we will discuss only numerical results in this case. For instance, for case II in Table~I depicted in Fig.~5, we obtain rotation periods $T =  5.0$ and $T=5.9$ for displacements $\Delta = 1$ and $\Delta = 2$. In dimensional units, for the parameter set II, this gives $T = 1.0 \times 10^1$~yr. and $T = 1.2 \times 10^1$~yr. for displacements $\Delta = 6.4 \times 10^{-2}$~pc and $\Delta = 1.3 \times 10^{-1}$ pc, respectively. We note that if BH induces a large gravielectric field on the  length scale of BEC localization, it significantly modifies the DM density profile (like in Fig.~5). This leads to a sufficient increase of the linear velocity $\omega \Delta$ of a vortex line. Vortex dynamics in this case requires an in-depth analysis which should incorporate relativistic effects whose comprehensive investigation lies beyond the scope of the present work.



\section{Conclusions}
\label{sec:conclusions}
We have investigated the dynamics of superfluid dark matter in the central regions of a typical spiral galaxy, where a spherically symmetric dark matter halo and a rotating supermassive black hole  coexist. In this scenario, dark matter forms a Bose-Einstein condensate core, behaving as a coherent quantum system. Using the Gross-Pitaevskii-Poisson model, we have studied the gravitational influence of the BH through both gravielectric and gravimagnetic fields.


Our analysis first focused on the stationary configurations of the BEC in the Kerr BH background. The gravielectric potential of the BH leads to formation of a pronounced density peak in the central region of the BEC, which we described both numerically and analytically using the Thomas-Fermi approximation.


We have studied main properties of the vortex lines in the self-gravitating BEC in the presence of the rotating BH. We have found that the rotation of the BH reduces the BEC energy through its gravimagnetic field, stabilizing the vortex lines under certain conditions. We have demonstrated the parameter regime where vortex lines become energetically favorable compared to the ground state.


The inhomogeneous background density defined by the total gravielectric potential of the BEC and BH system plays an important role in the dynamics of vortex lines. Numerical simulations show that the vortex line displaced from the BH rotation axis follows a circular trajectory around the rotation axis with constant angular velocity. This phenomenon, also known from the studies with atom BECs, is caused by the relative flow between the vortex and the ambient condensate and was analyzed analytically and numerically. 
We obtained analytical estimates for the precession frequency of a vortex line in self-gravitating BEC as well as BEC in the gravitational field of a supermassive BH.

Our findings clarify the role of black hole rotation on the quantum mechanical behavior of ultralight dark matter in galactic centers. The formation and precession of vortex lines, shaped by gravielectric and gravimagnetic fields, provide deeper insights into the manifestation of quantum effects on astrophysical scales. These results offer a promising framework for probing the interplay between general relativity and quantum theory in the context of dark matter, with potential observational signatures tied to vortex dynamics. A few interesting open questions are the BEC absorbtion by a BH and BH superradiance, which may be important but cannot be described within our model. We anticipate that this work will stimulate further investigations into the quantum nature of dark matter near rotating black holes, advancing our understanding of the galactic structure and dynamics.

\section{Acknowledgements}

We are grateful to Yuriy Bidasyuk for numerical code and useful discussions. The authors also acknowledge helpful discussions with Luca Salasnich.
K. K. acknowledges funding by the Deutsche Forschungsgemeinschaft under Germany’s Excellence Strategy EXC 2123 Quantum Frontiers Grant No. 390837967. A.I.Y. acknowledges support from the projects ‘Ultracold atoms in curved geometries’, 'Theoretical analysis of quantum atomic mixtures' of the University of Padova, and from INFN. 
O.O.P. acknowledges support from the National Research Foundation of Ukraine through Grant No. 2020.02/0032.

\appendix

\section{Kerr BH metric}
\label{App: Notes on spacetime metric}

The Kerr BH is a stationary axially symmetric solution described by the metric \cite{Wald}
\begin{multline*} 
     dS^{2}_\textrm{BH} = \left(1 - \frac{2GM_\textrm{BH}r}{c^{2}\rho^{2}}\right)(dx^{0})^{2} \\
     - \frac{2GM_\textrm{BH} ar \sin^{2}\theta}{c^{2}\rho^{2}}(dx^{0} d\phi + d\phi dx^{0}) - \frac{\rho^{2}}{\zeta}dr^{2} -\\- \rho^{2}d\theta^{2} - \frac{\sin^{2}\theta}{\rho^{2}}\left[(r^{2} + a^{2})^{2} - a^{2}\zeta \sin^{2}\theta\right]d\phi^{2},
\end{multline*}
where $\zeta(r) = r^{2} - \frac{2GM_{\textrm{BH}}}{c^{2}}r + a^{2}$ and $\rho^{2}(r, \theta) = r^{2} + a^{2}\cos^{2} \theta$. The Kerr BH metric is parametrized by two constants: the BH mass $M_\textrm{BH}$ and the rotation parameter $a = J/(cM_\textrm{BH})$. Observations of the central black hole of the Milky Way in Saggitarius A show that \cite{Melia_2001} $a \approx 0.19-0.34$ in units $R_\textrm{S}$ or $a \approx 0.99$ \cite{Aschenbach}. In general, $0 < a < 1$ in $R_\textrm{S}$ units \cite{Reynolds_2021}. Since $a \propto R_\textrm{S}$, both Schwarzschild radius and $a$ can be treated as small parameters in the regime $r \gg R_\textrm{S}$.
Then it is sufficient to keep only terms of the lowest order in both $R_\textrm{S}/r$ and $a/r$. In the first order, all terms with $a/r$ vanish and the metric reduces to the Schwarzshild form. Therefore, to describe the post-Newtonian effect of BH rotation, we need to include terms of the second order.

\section{Derivation of GP equation}
\label{App: Derivation of GP equation}

In the following we show how the non-relativistic model of GPP equations~(\ref{GP equation dimensionless}-\ref{Poisson equation dimensionless}) was derived from a more general relativistic model of a Klein-Gordon equation in curved spacetime
\begin{equation} 
\nabla_{\alpha}\nabla^{\alpha} \phi + \left[\left(\frac{mc}{\hbar}\right)^{2}  - U(|\phi|^{2}) \right]\phi = 0
\label{eq: KG equation}
\end{equation}
where $U = \frac{2m}{\hbar^{2}}gN|\phi|^{2}$ and $\phi$ is relativistic wave-function. Similar problem, but in the absence of BH was previously considered in \cite{korshynska2023dynamical}. Here $\nabla_{\alpha}$ denotes covariant derivative in the curved space-time. The sources of the space-time curvature are BH and gravitational field of the DM particles. Those in the limit of weak gravity, give independent contributions $\gamma^\textrm{DM}_{\mu \nu}$ and $\gamma^\textrm{BH}_{\mu \nu}$ to the linearized metric
\begin{equation*}
  ds^{2} = g_{\mu \nu}dx^{\mu}dx^{\nu} = (\eta_{\mu \nu} + \gamma_{\mu \nu}^\textrm{DM} + \gamma_{\mu \nu}^\textrm{BH})dx^{\mu}dx^{\nu},
\end{equation*}
explicit form of which is given by Eq.~(\ref{eq: spacetime metric}).

Metric Laplace operator of KG equation can be rewritten in the next way:
\begin{equation}
\nabla_{\alpha}\nabla^{\alpha}\phi = \frac{1}{\sqrt{-g}}\partial_{\alpha}(\sqrt{-g} g^{\alpha \beta} \partial_{\beta}\phi),
\label{eq: Metric Laplace operator}
\end{equation}
where $g = \textrm{det}(g_{\mu \nu})$. The explicit metric expression yields the determinant
\begin{multline*}
  g  = -r^{4}\sin^{2}\theta \left(1 - \frac{4\Phi^\textrm{DM}}{c^{2}} + \frac{2a^{2}\cos^{2} \theta}{r^{2}}\right) \\ + O\left[\left(\frac{\Phi^\textrm{DM}}{c^{2}}\right)^{2}, \left(\frac{R_\textrm{S}}{r}\right)^{3}, \frac{R_\textrm{S}}{r}\frac{\Phi^\textrm{DM}}{c^{2}}\right]
\end{multline*}

Then one can substitute this result into relation (\ref{eq: Metric Laplace operator}) together with the explicit metric components~(\ref{eq: spacetime metric}) and get the relativistic equation for the DM in the curved spacetime. We rewrite this equation in the non-relativistic limit with $\Psi = e^{-imc^{2}t/\hbar}\psi$, where $\psi$ changes slowly with time. After neglecting small terms 
$\propto \frac{\partial \psi}{\partial t}\frac{\Phi}{c^{2}}$ and $\propto \Phi \Delta \psi$ this gives
\begin{multline*}
    i\hbar \frac{\partial \psi}{\partial t}  = -\frac{\hbar^{2}}{2m}\Delta \psi \\- \frac{mc^{2}}{2}\left[\frac{2R_\textrm{S}}{r} + \left(\frac{2R_\textrm{S}}{r}\right)^{2} - \frac{2\Phi^\textrm{DM}}{c^{2}}\right]\psi \\- 2\frac{i\hbar c}{r}\left[-\frac{R_\textrm{S}a}{r^{2}} + \frac{A_{\phi}^\textrm{DM}}{c^{2}\sin \theta}\right]\frac{\partial \psi}{\partial \phi}  - U(|\psi|^{2})\psi ,
\end{multline*}
where we kept only the terms in the highest linear order in $a$.

In the paper we introduced total gravimagnetic potential $\mathbf{A} = A_{\phi}\mathbf{e}_{\phi} = \left(A_{\phi}^\textrm{DM} - c^{2}\frac{aR_\textrm{S}}{r^{2}}\sin \theta\right)$ and total gravielectric potential $\Phi = \Phi^\textrm{DM} - c^{2}\frac{R_\textrm{S}}{r}$. The latter in KG equation is modified and reads $\tilde{\Phi} = \Phi - \frac{c^{2}}{2}\left(\frac{2R_\textrm{S}}{r}\right)^{2}$. Using this notations, we finally obtain Eq.~(\ref{GP equation dimensionless}).

\section{Dark matter gravimagnetic field and soliton absorption time}
\label{Appendix: Model restrictions}

In our model given by Eqs.~(\ref{modeleq1})-(\ref{modeleq2}), we assume that the gravimagnetic field $\mathbf{A}_\textrm{DM}$ induced by DM is  much smaller compared to the gravimagnetic field $\mathbf{A}_\textrm{BH}$ induced by BH. Let us estimate the magnitudes of these fields assuming that the vortex is situated at $\mathbf{r}_0$ with respect to BH. The gravimagnetic field couples to the BEC current, thus, it affects BEC on the scale of the vortex width, i.e., for $|\mathbf{r} - \mathbf{r}_0| < \xi$.

 The gravimagnetic potential of BH is of order $A_\textrm{BH} \approx c^2 R_\textrm{s}^2/r_0^2$ at this distance.  On the other hand, the DM gravimagnetic potential can be estimated as 
\begin{equation*}
    A^\textrm{DM} \propto \frac{G}{c} \rho v L^2,
\end{equation*}
where $\rho$ and $v$ are local values of the DM density and velocity, respectively. Using Eq.~(\ref{eq: vortex line ansatz}) we see that $\rho \propto (|\mathbf{r} - \mathbf{r}_0|/\xi)^2 |\psi_0(\Delta)|^2 \approx (|\mathbf{r} - \mathbf{r}_0|/\xi)^2 \times M_\textrm{DM}/(4 \pi^2 L^3)$ and $v \propto c\lambda_\textrm{c}/L \times L/|\mathbf{r} - \mathbf{r}_0|$ in the plane perpendicular to the vortex axis. Then we get
\begin{equation*}
    A^\textrm{DM} \propto G\frac{M_\textrm{DM}}{4\pi^2} \frac{|\mathbf{r} - \mathbf{r}_0|}{\xi^2} \frac{\lambda_\textrm{c}}{L} \propto \frac{G M_\textrm{DM} }{4\pi^2 \xi} \frac{\lambda_\textrm{c}}{L},
\end{equation*}
and, therefore,
\begin{equation*}
   \frac{A^\textrm{DM}}{A^\textrm{BH}} \propto \frac{G M_\textrm{DM}}{4 \pi^2 L c^2} \left(\frac{\xi}{L} \right)^{-1}\frac{\lambda_\textrm{c}}{L} \left(\frac{R_\textrm{s}}{L}\right)^{-2} \left(\frac{L}{r_0}\right)^2.
\end{equation*}

This formula can be further simplified to
\begin{equation}
\frac{A^\textrm{DM}}{A^\textrm{BH}} \propto 10 \left(\frac{\lambda_\textrm{c}}{L}\right)^3\left(\frac{R_\textrm{s}}{L}\right)^{-2} \left(\frac{L}{r_0}\right)^2
\end{equation}
assuming $N_0 = 10^3$ (here we used relations (\ref{eq: DM mass}) - (\ref{eq: DM radius})). From here we can see that it is not always the case that $A^\textrm{DM} < A^\textrm{BH}$. In fact, for a parameter set I (see Table~I) DM provides the dominating contribution to the gravimagnetic field, though in this case both $A^\textrm{DM}$ and $A^\textrm{BH}$ are small and thus neglected. In general, the only effect of gravimagnetic field is the decrease of vortex energy, discussed in Sec.~\ref{subsec: Energy analysis}, where the presence of $A^\textrm{DM}$ indeed does not sufficiently affect our calculations.

As we mentioned in the Introduction, our model is applicable only at time scales $t \ll \tau$, where $\tau$ is the characteristic time of the DM soliton absorption by a supermassive BH. According to \cite{bar2018galactic}, the BH absorption time in the model of non-interacting self-gravitating ULDM is characterized by the parameter
\begin{equation}
    \zeta = 0.04 \left(\frac{m}{10^{-22} \textrm{eV}}\right) \left(\frac{M_\textrm{BH}}{10^6 M_\odot}\right)\left(\frac{M_\textrm{DM}}{10^{12} M_\odot}\right)^{-1/3}.\label{eq: zeta}
\end{equation}
If  $\zeta \ll 1$, then the absorption time equals
\begin{multline}
    \frac{\tau}{1 \textrm{y} } \propto 2.4 \times 10^{17} \left(\frac{m}{10^{-22} \textrm{eV}}\right)^{-2} \\ \times\left(\frac{M_\textrm{BH}}{4 \times 10^6 M_\odot}\right)^{-1}\left(\frac{M_\textrm{DM}}{10^{12} M_\odot}\right)^{-4/3}.
\end{multline}
For $\zeta \gg 1$, the absorption time is given by 
\begin{multline}
    \frac{\tau}{1 \textrm{y}} \propto 1.5 \times 10^{18} \left(\frac{m}{10^{-22} \textrm{eV}}\right)^{-3} \\ \times\left(\frac{M_\textrm{BH}}{4 \times 10^6 M_\odot}\right)^{-2}\left(\frac{M_\textrm{DM}}{10^{12} M_\odot}\right)^{-1}.\label{eq: accretion rate}
\end{multline}
Clearly, this time is much larger than the Hubble time in the case of the supermassive Sgr A* BH in the Milky Way and the bosonic particle mass $m = 10^{-21}$ eV chosen in this study.

According to Eq.~(\ref{eq: zeta}), we have $\zeta \gg 1$ in both cases I and II present in Table~I. Then, in view of Eq.(\ref{eq: accretion rate}), the absorption time  equals $\tau = 1.34 \times 10^{14}\, \textrm{yr.} \gg \omega_*^{-1}$ for case I and $\tau = 1.34 \times 10^6\,\textrm{yr.} \gg \omega_*^{-1}$ for case II. Thus, for the chosen two sets of parameters, the effect of BEC absorption by BH can be neglected.

\section{Dynamics of vortex line}
\label{App: Dynamics of a vortex line}

There are a few techniques to analytically estimate the frequency of vortex line precession around the trap center discussed in \cite{fetter2001vortices, groszek2018motion}. Following notations in the main text, we denote vortex position at the moment of time $t$ as $\mathbf{r}_0 = \{x_0, y_0\}$. For our estimate, we will use the fact that the energy variation with respect to $\mathbf{r}_0$ should be balanced by the Magnus force (see, \cite{jackson1999vortex, fetter2001vortices})
\begin{equation*}
    \mathbf{F}_\textrm{Mag} = \frac{\partial E(\mathbf{r}_0)}{\partial \mathbf{r}_0} = m \rho (\mathbf{r}_0) [\mathbf{\kappa} \times \mathbf{v}_0].
\end{equation*}
Here $\rho(\mathbf{r}_0)$ stands for the TF ground state number density at the vortex location, $m$ is the boson mass, and $\mathbf{\kappa} = h/m \mathbf{e}_z$ is a circulation vector. 

Using the axial symmetry of the system we introduce polar coordinates $r_0 = \sqrt{x_0^2 + y_0^2}$ and $\phi_0 = \arctan(y_0/x_0)$, so that the energy $E(\mathbf{r}_0) = E(r_0)$ is a function of radial vortex position only. In this case the above equation leads to a fixed radius trajectory $dr_0/dt = 0$ with the precession frequency
\begin{equation*}
    \frac{d\phi_0}{dt} = - \frac{1}{h r_0 \rho(r_0)}\frac{\partial E}{\partial r_0}.
\end{equation*}

Following the same argument as presented in the context of an atom BEC in a harmonic trap \cite{jackson1999vortex}, we note that the leading contribution to the $\partial E/\partial r_0$ is given by the nonlinear term $\propto |\psi|^2$ in Eq.~(\ref{eq: energy}). Moreover, we assume that the effect of the vortex line deformation can be neglected, i.e., the vortex line moves as a whole. Therefore, the vortex line dynamics can be described by the motion of a point-like vortex in the 2D plane $z=0$ with frequency
\begin{equation*}
    \frac{d\phi_0}{dt} = - \frac{g}{2 h r_0 |\psi_0(r_0)|^2}\frac{\partial }{\partial r_0}\left(\int d^2\mathbf{r} |\psi|^4\right).
\end{equation*}
For consistency, we convert the expression above to the dimensionless units introduced in Sec.~\ref{sec:ansatz} and get
\begin{equation*}
    \omega_0 = - \frac{1}{4 \pi r_0 |\psi_0(r_0)|^2}\frac{\partial}{\partial r_0}\left(\int d^2\mathbf{r} |\psi|^4\right),
\end{equation*}
where $\omega_0$ is now measured in units $\omega_*$.

\bibliography{Bibl}

\begin{thebibliography}{60}%
\makeatletter
\providecommand \@ifxundefined [1]{%
 \@ifx{#1\undefined}
}%
\providecommand \@ifnum [1]{%
 \ifnum #1\expandafter \@firstoftwo
 \else \expandafter \@secondoftwo
 \fi
}%
\providecommand \@ifx [1]{%
 \ifx #1\expandafter \@firstoftwo
 \else \expandafter \@secondoftwo
 \fi
}%
\providecommand \natexlab [1]{#1}%
\providecommand \enquote  [1]{``#1''}%
\providecommand \bibnamefont  [1]{#1}%
\providecommand \bibfnamefont [1]{#1}%
\providecommand \citenamefont [1]{#1}%
\providecommand \href@noop [0]{\@secondoftwo}%
\providecommand \href [0]{\begingroup \@sanitize@url \@href}%
\providecommand \@href[1]{\@@startlink{#1}\@@href}%
\providecommand \@@href[1]{\endgroup#1\@@endlink}%
\providecommand \@sanitize@url [0]{\catcode `\\12\catcode `\$12\catcode `\&12\catcode `\#12\catcode `\^12\catcode `\_12\catcode `\%12\relax}%
\providecommand \@@startlink[1]{}%
\providecommand \@@endlink[0]{}%
\providecommand \url  [0]{\begingroup\@sanitize@url \@url }%
\providecommand \@url [1]{\endgroup\@href {#1}{\urlprefix }}%
\providecommand \urlprefix  [0]{URL }%
\providecommand \Eprint [0]{\href }%
\providecommand \doibase [0]{https://doi.org/}%
\providecommand \selectlanguage [0]{\@gobble}%
\providecommand \bibinfo  [0]{\@secondoftwo}%
\providecommand \bibfield  [0]{\@secondoftwo}%
\providecommand \translation [1]{[#1]}%
\providecommand \BibitemOpen [0]{}%
\providecommand \bibitemStop [0]{}%
\providecommand \bibitemNoStop [0]{.\EOS\space}%
\providecommand \EOS [0]{\spacefactor3000\relax}%
\providecommand \BibitemShut  [1]{\csname bibitem#1\endcsname}%
\let\auto@bib@innerbib\@empty
\bibitem [{\citenamefont {Ferreira}(2021)}]{Ferreira}%
  \BibitemOpen
  \bibfield  {author} {\bibinfo {author} {\bibfnamefont {E.~G.~M.}\ \bibnamefont {Ferreira}},\ }\bibfield  {journal} {\bibinfo  {journal} {The Astronomy and Astrophysics Review}\ }\textbf {\bibinfo {volume} {29}},\ \href {https://doi.org/10.1007/s00159-021-00135-6} {10.1007/s00159-021-00135-6} (\bibinfo {year} {2021})\BibitemShut {NoStop}%
\bibitem [{\citenamefont {Jackson~Kimball}\ and\ \citenamefont {Van~Bibber}(2023)}]{jackson2023search}%
  \BibitemOpen
  \bibfield  {author} {\bibinfo {author} {\bibfnamefont {D.~F.}\ \bibnamefont {Jackson~Kimball}}\ and\ \bibinfo {author} {\bibfnamefont {K.}~\bibnamefont {Van~Bibber}},\ }\href@noop {} {\emph {\bibinfo {title} {The search for ultralight bosonic dark matter}}}\ (\bibinfo  {publisher} {Springer Nature},\ \bibinfo {year} {2023})\BibitemShut {NoStop}%
\bibitem [{\citenamefont {Böhmer}\ and\ \citenamefont {Harko}(2007)}]{B_hmer_2007}%
  \BibitemOpen
  \bibfield  {author} {\bibinfo {author} {\bibfnamefont {C.~G.}\ \bibnamefont {Böhmer}}\ and\ \bibinfo {author} {\bibfnamefont {T.}~\bibnamefont {Harko}},\ }\href {https://doi.org/10.1088/1475-7516/2007/06/025} {\bibfield  {journal} {\bibinfo  {journal} {Journal of Cosmology and Astroparticle Physics}\ }\textbf {\bibinfo {volume} {2007}}\bibinfo  {number} { (06)},\ \bibinfo {pages} {025}}\BibitemShut {NoStop}%
\bibitem [{\citenamefont {Chavanis}(2016)}]{https://doi.org/10.48550/arxiv.1611.09610}%
  \BibitemOpen
\bibfield  {number} {  }\bibfield  {author} {\bibinfo {author} {\bibfnamefont {P.-H.}\ \bibnamefont {Chavanis}},\ }\href {https://doi.org/10.1140/epjp/i2017-11544-3} {\bibfield  {journal} {\bibinfo  {journal} {European Physical Journal Plus}\ }\textbf {\bibinfo {volume} {132}} (\bibinfo {year} {2016})}\BibitemShut {NoStop}%
\bibitem [{\citenamefont {Chavanis}\ and\ \citenamefont {Harko}(2012)}]{PhysRevD.86.064011}%
  \BibitemOpen
  \bibfield  {author} {\bibinfo {author} {\bibfnamefont {P.-H.}\ \bibnamefont {Chavanis}}\ and\ \bibinfo {author} {\bibfnamefont {T.}~\bibnamefont {Harko}},\ }\href {https://doi.org/10.1103/PhysRevD.86.064011} {\bibfield  {journal} {\bibinfo  {journal} {Phys. Rev. D}\ }\textbf {\bibinfo {volume} {86}},\ \bibinfo {pages} {064011} (\bibinfo {year} {2012})}\BibitemShut {NoStop}%
\bibitem [{\citenamefont {Hui}\ \emph {et~al.}(2017)\citenamefont {Hui}, \citenamefont {Ostriker}, \citenamefont {Tremaine},\ and\ \citenamefont {Witten}}]{Hui_2017}%
  \BibitemOpen
  \bibfield  {author} {\bibinfo {author} {\bibfnamefont {L.}~\bibnamefont {Hui}}, \bibinfo {author} {\bibfnamefont {J.~P.}\ \bibnamefont {Ostriker}}, \bibinfo {author} {\bibfnamefont {S.}~\bibnamefont {Tremaine}},\ and\ \bibinfo {author} {\bibfnamefont {E.}~\bibnamefont {Witten}},\ }\bibfield  {journal} {\bibinfo  {journal} {Physical Review D}\ }\textbf {\bibinfo {volume} {95}},\ \href {https://doi.org/10.1103/physrevd.95.043541} {10.1103/physrevd.95.043541} (\bibinfo {year} {2017})\BibitemShut {NoStop}%
\bibitem [{\citenamefont {Rindler-Daller}\ and\ \citenamefont {Shapiro}(2014)}]{rindler2014complex}%
  \BibitemOpen
  \bibfield  {author} {\bibinfo {author} {\bibfnamefont {T.}~\bibnamefont {Rindler-Daller}}\ and\ \bibinfo {author} {\bibfnamefont {P.~R.}\ \bibnamefont {Shapiro}},\ }\href@noop {} {\bibfield  {journal} {\bibinfo  {journal} {Modern Physics Letters A}\ }\textbf {\bibinfo {volume} {29}},\ \bibinfo {pages} {1430002} (\bibinfo {year} {2014})}\BibitemShut {NoStop}%
\bibitem [{\citenamefont {Sikivie}\ and\ \citenamefont {Yang}(2009)}]{sikivie2009bose}%
  \BibitemOpen
  \bibfield  {author} {\bibinfo {author} {\bibfnamefont {P.}~\bibnamefont {Sikivie}}\ and\ \bibinfo {author} {\bibfnamefont {Q.}~\bibnamefont {Yang}},\ }\href@noop {} {\bibfield  {journal} {\bibinfo  {journal} {Physical Review Letters}\ }\textbf {\bibinfo {volume} {103}},\ \bibinfo {pages} {111301} (\bibinfo {year} {2009})}\BibitemShut {NoStop}%
\bibitem [{\citenamefont {Schive}\ \emph {et~al.}(2014{\natexlab{a}})\citenamefont {Schive}, \citenamefont {Chiueh},\ and\ \citenamefont {Broadhurst}}]{Schive_2014}%
  \BibitemOpen
  \bibfield  {author} {\bibinfo {author} {\bibfnamefont {H.-Y.}\ \bibnamefont {Schive}}, \bibinfo {author} {\bibfnamefont {T.}~\bibnamefont {Chiueh}},\ and\ \bibinfo {author} {\bibfnamefont {T.}~\bibnamefont {Broadhurst}},\ }\href {https://doi.org/10.1038/nphys2996} {\bibfield  {journal} {\bibinfo  {journal} {Nature Physics}\ }\textbf {\bibinfo {volume} {10}},\ \bibinfo {pages} {496} (\bibinfo {year} {2014}{\natexlab{a}})}\BibitemShut {NoStop}%
\bibitem [{\citenamefont {Matos}\ and\ \citenamefont {Ure{\~{n} }a-L{\'{o}}pez}(2000)}]{Matos_2000}%
  \BibitemOpen
  \bibfield  {author} {\bibinfo {author} {\bibfnamefont {T.}~\bibnamefont {Matos}}\ and\ \bibinfo {author} {\bibfnamefont {L.~A.}\ \bibnamefont {Ure{\~{n} }a-L{\'{o}}pez}},\ }\href {https://doi.org/10.1088/0264-9381/17/13/101} {\bibfield  {journal} {\bibinfo  {journal} {Classical and Quantum Gravity}\ }\textbf {\bibinfo {volume} {17}},\ \bibinfo {pages} {L75} (\bibinfo {year} {2000})}\BibitemShut {NoStop}%
\bibitem [{\citenamefont {Sahni}\ and\ \citenamefont {Wang}(2000)}]{PhysRevD.62.103517}%
  \BibitemOpen
  \bibfield  {author} {\bibinfo {author} {\bibfnamefont {V.}~\bibnamefont {Sahni}}\ and\ \bibinfo {author} {\bibfnamefont {L.}~\bibnamefont {Wang}},\ }\href {https://doi.org/10.1103/PhysRevD.62.103517} {\bibfield  {journal} {\bibinfo  {journal} {Phys. Rev. D}\ }\textbf {\bibinfo {volume} {62}},\ \bibinfo {pages} {103517} (\bibinfo {year} {2000})}\BibitemShut {NoStop}%
\bibitem [{\citenamefont {Bar}\ \emph {et~al.}(2018)\citenamefont {Bar}, \citenamefont {Blas}, \citenamefont {Blum},\ and\ \citenamefont {Sibiryakov}}]{bar2018galactic}%
  \BibitemOpen
  \bibfield  {author} {\bibinfo {author} {\bibfnamefont {N.}~\bibnamefont {Bar}}, \bibinfo {author} {\bibfnamefont {D.}~\bibnamefont {Blas}}, \bibinfo {author} {\bibfnamefont {K.}~\bibnamefont {Blum}},\ and\ \bibinfo {author} {\bibfnamefont {S.}~\bibnamefont {Sibiryakov}},\ }\href@noop {} {\bibfield  {journal} {\bibinfo  {journal} {Physical Review D}\ }\textbf {\bibinfo {volume} {98}},\ \bibinfo {pages} {083027} (\bibinfo {year} {2018})}\BibitemShut {NoStop}%
\bibitem [{\citenamefont {De~Martino}\ \emph {et~al.}(2020)\citenamefont {De~Martino}, \citenamefont {Broadhurst}, \citenamefont {Tye}, \citenamefont {Chiueh},\ and\ \citenamefont {Schive}}]{de2020dynamical}%
  \BibitemOpen
  \bibfield  {author} {\bibinfo {author} {\bibfnamefont {I.}~\bibnamefont {De~Martino}}, \bibinfo {author} {\bibfnamefont {T.}~\bibnamefont {Broadhurst}}, \bibinfo {author} {\bibfnamefont {S.-H.~H.}\ \bibnamefont {Tye}}, \bibinfo {author} {\bibfnamefont {T.}~\bibnamefont {Chiueh}},\ and\ \bibinfo {author} {\bibfnamefont {H.-Y.}\ \bibnamefont {Schive}},\ }\href@noop {} {\bibfield  {journal} {\bibinfo  {journal} {Physics of the Dark Universe}\ }\textbf {\bibinfo {volume} {28}},\ \bibinfo {pages} {100503} (\bibinfo {year} {2020})}\BibitemShut {NoStop}%
\bibitem [{\citenamefont {Goldstein}\ \emph {et~al.}(2022)\citenamefont {Goldstein}, \citenamefont {Koushiappas},\ and\ \citenamefont {Walker}}]{goldstein2022viability}%
  \BibitemOpen
  \bibfield  {author} {\bibinfo {author} {\bibfnamefont {I.~S.}\ \bibnamefont {Goldstein}}, \bibinfo {author} {\bibfnamefont {S.~M.}\ \bibnamefont {Koushiappas}},\ and\ \bibinfo {author} {\bibfnamefont {M.~G.}\ \bibnamefont {Walker}},\ }\href@noop {} {\bibfield  {journal} {\bibinfo  {journal} {Physical Review D}\ }\textbf {\bibinfo {volume} {106}},\ \bibinfo {pages} {063010} (\bibinfo {year} {2022})}\BibitemShut {NoStop}%
\bibitem [{\citenamefont {Gan}\ \emph {et~al.}(2024)\citenamefont {Gan}, \citenamefont {Wang},\ and\ \citenamefont {Xiao}}]{gan2024detecting}%
  \BibitemOpen
  \bibfield  {author} {\bibinfo {author} {\bibfnamefont {X.}~\bibnamefont {Gan}}, \bibinfo {author} {\bibfnamefont {L.-T.}\ \bibnamefont {Wang}},\ and\ \bibinfo {author} {\bibfnamefont {H.}~\bibnamefont {Xiao}},\ }\href@noop {} {\bibfield  {journal} {\bibinfo  {journal} {Physical Review D}\ }\textbf {\bibinfo {volume} {110}},\ \bibinfo {pages} {063039} (\bibinfo {year} {2024})}\BibitemShut {NoStop}%
\bibitem [{\citenamefont {Harvey}\ \emph {et~al.}(2015)\citenamefont {Harvey}, \citenamefont {Massey}, \citenamefont {Kitching}, \citenamefont {Taylor},\ and\ \citenamefont {Tittley}}]{Harvey_2015}%
  \BibitemOpen
  \bibfield  {author} {\bibinfo {author} {\bibfnamefont {D.}~\bibnamefont {Harvey}}, \bibinfo {author} {\bibfnamefont {R.}~\bibnamefont {Massey}}, \bibinfo {author} {\bibfnamefont {T.}~\bibnamefont {Kitching}}, \bibinfo {author} {\bibfnamefont {A.}~\bibnamefont {Taylor}},\ and\ \bibinfo {author} {\bibfnamefont {E.}~\bibnamefont {Tittley}},\ }\href {https://doi.org/10.1126/science.1261381} {\bibfield  {journal} {\bibinfo  {journal} {Science}\ }\textbf {\bibinfo {volume} {347}},\ \bibinfo {pages} {1462} (\bibinfo {year} {2015})}\BibitemShut {NoStop}%
\bibitem [{\citenamefont {{Lee}}\ \emph {et~al.}(2008)\citenamefont {{Lee}}, \citenamefont {{Lim}},\ and\ \citenamefont {{Choi}}}]{2008arXiv0805.3827L}%
  \BibitemOpen
  \bibfield  {author} {\bibinfo {author} {\bibfnamefont {J.-W.}\ \bibnamefont {{Lee}}}, \bibinfo {author} {\bibfnamefont {S.}~\bibnamefont {{Lim}}},\ and\ \bibinfo {author} {\bibfnamefont {D.}~\bibnamefont {{Choi}}},\ }\href@noop {} {\bibfield  {journal} {\bibinfo  {journal} {arXiv e-prints}\ ,\ \bibinfo {eid} {arXiv:0805.3827}} (\bibinfo {year} {2008})},\ \Eprint {https://arxiv.org/abs/0805.3827} {arXiv:0805.3827 [hep-ph]} \BibitemShut {NoStop}%
\bibitem [{\citenamefont {Baryakhtar}\ \emph {et~al.}(2021)\citenamefont {Baryakhtar}, \citenamefont {Galanis}, \citenamefont {Lasenby},\ and\ \citenamefont {Simon}}]{baryakhtar2021black}%
  \BibitemOpen
  \bibfield  {author} {\bibinfo {author} {\bibfnamefont {M.}~\bibnamefont {Baryakhtar}}, \bibinfo {author} {\bibfnamefont {M.}~\bibnamefont {Galanis}}, \bibinfo {author} {\bibfnamefont {R.}~\bibnamefont {Lasenby}},\ and\ \bibinfo {author} {\bibfnamefont {O.}~\bibnamefont {Simon}},\ }\href@noop {} {\bibfield  {journal} {\bibinfo  {journal} {Physical Review D}\ }\textbf {\bibinfo {volume} {103}},\ \bibinfo {pages} {095019} (\bibinfo {year} {2021})}\BibitemShut {NoStop}%
\bibitem [{\citenamefont {Collaviti}\ \emph {et~al.}(2024)\citenamefont {Collaviti}, \citenamefont {Sun}, \citenamefont {Galanis},\ and\ \citenamefont {Baryakhtar}}]{collaviti2024observational}%
  \BibitemOpen
  \bibfield  {author} {\bibinfo {author} {\bibfnamefont {S.}~\bibnamefont {Collaviti}}, \bibinfo {author} {\bibfnamefont {L.}~\bibnamefont {Sun}}, \bibinfo {author} {\bibfnamefont {M.}~\bibnamefont {Galanis}},\ and\ \bibinfo {author} {\bibfnamefont {M.}~\bibnamefont {Baryakhtar}},\ }\href@noop {} {\bibfield  {journal} {\bibinfo  {journal} {arXiv preprint arXiv:2407.04304}\ } (\bibinfo {year} {2024})}\BibitemShut {NoStop}%
\bibitem [{\citenamefont {Armengaud}(2019)}]{armengaud2019supermassive}%
  \BibitemOpen
  \bibfield  {author} {\bibinfo {author} {\bibfnamefont {E.}~\bibnamefont {Armengaud}},\ }\href@noop {} {\bibfield  {journal} {\bibinfo  {journal} {Physics}\ }\textbf {\bibinfo {volume} {12}},\ \bibinfo {pages} {78} (\bibinfo {year} {2019})}\BibitemShut {NoStop}%
\bibitem [{\citenamefont {Chavanis}(2019{\natexlab{a}})}]{chavanis2019predictive}%
  \BibitemOpen
  \bibfield  {author} {\bibinfo {author} {\bibfnamefont {P.-H.}\ \bibnamefont {Chavanis}},\ }\href@noop {} {\bibfield  {journal} {\bibinfo  {journal} {Physical Review D}\ }\textbf {\bibinfo {volume} {100}},\ \bibinfo {pages} {083022} (\bibinfo {year} {2019}{\natexlab{a}})}\BibitemShut {NoStop}%
\bibitem [{\citenamefont {Chavanis}(2022)}]{chavanis2022heuristic}%
  \BibitemOpen
  \bibfield  {author} {\bibinfo {author} {\bibfnamefont {P.-H.}\ \bibnamefont {Chavanis}},\ }\href@noop {} {\bibfield  {journal} {\bibinfo  {journal} {The European Physical Journal B}\ }\textbf {\bibinfo {volume} {95}},\ \bibinfo {pages} {48} (\bibinfo {year} {2022})}\BibitemShut {NoStop}%
\bibitem [{\citenamefont {Launhardt}\ \emph {et~al.}(2002)\citenamefont {Launhardt}, \citenamefont {Zylka},\ and\ \citenamefont {Mezger}}]{launhardt2002nuclear}%
  \BibitemOpen
  \bibfield  {author} {\bibinfo {author} {\bibfnamefont {R.}~\bibnamefont {Launhardt}}, \bibinfo {author} {\bibfnamefont {R.}~\bibnamefont {Zylka}},\ and\ \bibinfo {author} {\bibfnamefont {P.}~\bibnamefont {Mezger}},\ }\href@noop {} {\bibfield  {journal} {\bibinfo  {journal} {Astronomy $\&$ Astrophysics}\ }\textbf {\bibinfo {volume} {384}},\ \bibinfo {pages} {112} (\bibinfo {year} {2002})}\BibitemShut {NoStop}%
\bibitem [{\citenamefont {Sch{\"o}nrich}\ \emph {et~al.}(2015)\citenamefont {Sch{\"o}nrich}, \citenamefont {Aumer},\ and\ \citenamefont {Sale}}]{schonrich2015kinematic}%
  \BibitemOpen
  \bibfield  {author} {\bibinfo {author} {\bibfnamefont {R.}~\bibnamefont {Sch{\"o}nrich}}, \bibinfo {author} {\bibfnamefont {M.}~\bibnamefont {Aumer}},\ and\ \bibinfo {author} {\bibfnamefont {S.~E.}\ \bibnamefont {Sale}},\ }\href@noop {} {\bibfield  {journal} {\bibinfo  {journal} {The Astrophysical Journal Letters}\ }\textbf {\bibinfo {volume} {812}},\ \bibinfo {pages} {L21} (\bibinfo {year} {2015})}\BibitemShut {NoStop}%
\bibitem [{\citenamefont {Portail}\ \emph {et~al.}(2016)\citenamefont {Portail}, \citenamefont {Gerhard}, \citenamefont {Wegg},\ and\ \citenamefont {Ness}}]{portail2016dynamical}%
  \BibitemOpen
  \bibfield  {author} {\bibinfo {author} {\bibfnamefont {M.}~\bibnamefont {Portail}}, \bibinfo {author} {\bibfnamefont {O.}~\bibnamefont {Gerhard}}, \bibinfo {author} {\bibfnamefont {C.}~\bibnamefont {Wegg}},\ and\ \bibinfo {author} {\bibfnamefont {M.}~\bibnamefont {Ness}},\ }\href@noop {} {\bibfield  {journal} {\bibinfo  {journal} {Monthly Notices of the Royal Astronomical Society}\ ,\ \bibinfo {pages} {stw2819}} (\bibinfo {year} {2016})}\BibitemShut {NoStop}%
\bibitem [{\citenamefont {Schive}\ \emph {et~al.}(2014{\natexlab{b}})\citenamefont {Schive}, \citenamefont {Liao}, \citenamefont {Woo}, \citenamefont {Wong}, \citenamefont {Chiueh}, \citenamefont {Broadhurst},\ and\ \citenamefont {Hwang}}]{Schive_2014-core-halo}%
  \BibitemOpen
  \bibfield  {author} {\bibinfo {author} {\bibfnamefont {H.-Y.}\ \bibnamefont {Schive}}, \bibinfo {author} {\bibfnamefont {M.-H.}\ \bibnamefont {Liao}}, \bibinfo {author} {\bibfnamefont {T.-P.}\ \bibnamefont {Woo}}, \bibinfo {author} {\bibfnamefont {S.-K.}\ \bibnamefont {Wong}}, \bibinfo {author} {\bibfnamefont {T.}~\bibnamefont {Chiueh}}, \bibinfo {author} {\bibfnamefont {T.}~\bibnamefont {Broadhurst}},\ and\ \bibinfo {author} {\bibfnamefont {W.-Y.~P.}\ \bibnamefont {Hwang}},\ }\bibfield  {journal} {\bibinfo  {journal} {Physical Review Letters}\ }\textbf {\bibinfo {volume} {113}},\ \href {https://doi.org/10.1103/physrevlett.113.261302} {10.1103/physrevlett.113.261302} (\bibinfo {year} {2014}{\natexlab{b}})\BibitemShut {NoStop}%
\bibitem [{\citenamefont {Abuter}\ \emph {et~al.}(2023)\citenamefont {Abuter}, \citenamefont {Aimar}, \citenamefont {Seoane}, \citenamefont {Amorim}, \citenamefont {Baub{\"o}ck}, \citenamefont {Berger}, \citenamefont {Bonnet}, \citenamefont {Bourdarot}, \citenamefont {Brandner}, \citenamefont {Cardoso} \emph {et~al.}}]{abuter2023polarimetry}%
  \BibitemOpen
  \bibfield  {author} {\bibinfo {author} {\bibfnamefont {R.}~\bibnamefont {Abuter}}, \bibinfo {author} {\bibfnamefont {N.}~\bibnamefont {Aimar}}, \bibinfo {author} {\bibfnamefont {P.~A.}\ \bibnamefont {Seoane}}, \bibinfo {author} {\bibfnamefont {A.}~\bibnamefont {Amorim}}, \bibinfo {author} {\bibfnamefont {M.}~\bibnamefont {Baub{\"o}ck}}, \bibinfo {author} {\bibfnamefont {J.}~\bibnamefont {Berger}}, \bibinfo {author} {\bibfnamefont {H.}~\bibnamefont {Bonnet}}, \bibinfo {author} {\bibfnamefont {G.}~\bibnamefont {Bourdarot}}, \bibinfo {author} {\bibfnamefont {W.}~\bibnamefont {Brandner}}, \bibinfo {author} {\bibfnamefont {V.}~\bibnamefont {Cardoso}}, \emph {et~al.},\ }\href@noop {} {\bibfield  {journal} {\bibinfo  {journal} {arXiv preprint arXiv:2307.11821}\ } (\bibinfo {year} {2023})}\BibitemShut {NoStop}%
\bibitem [{\citenamefont {Merritt}(2013)}]{merritt2013dynamics}%
  \BibitemOpen
  \bibfield  {author} {\bibinfo {author} {\bibfnamefont {D.}~\bibnamefont {Merritt}},\ }\href@noop {} {\emph {\bibinfo {title} {Dynamics and evolution of galactic nuclei}}}\ (\bibinfo  {publisher} {Princeton University Press},\ \bibinfo {year} {2013})\BibitemShut {NoStop}%
\bibitem [{\citenamefont {Chavanis}(2019{\natexlab{b}})}]{chavanis2019mass}%
  \BibitemOpen
  \bibfield  {author} {\bibinfo {author} {\bibfnamefont {P.-H.}\ \bibnamefont {Chavanis}},\ }\href@noop {} {\bibfield  {journal} {\bibinfo  {journal} {The European Physical Journal Plus}\ }\textbf {\bibinfo {volume} {134}},\ \bibinfo {pages} {352} (\bibinfo {year} {2019}{\natexlab{b}})}\BibitemShut {NoStop}%
\bibitem [{\citenamefont {Chavanis}(2015)}]{Chavanis}%
  \BibitemOpen
  \bibfield  {author} {\bibinfo {author} {\bibfnamefont {P.-H.}\ \bibnamefont {Chavanis}},\ }\href {https://doi.org/10.1007/978-3-319-10852-0_6} {\bibfield  {journal} {\bibinfo  {journal} {Quantum Aspects of Black Holes}\ ,\ \bibinfo {pages} {151}} (\bibinfo {year} {2015})}\BibitemShut {NoStop}%
\bibitem [{\citenamefont {Hui}\ \emph {et~al.}(2021)\citenamefont {Hui}, \citenamefont {Joyce}, \citenamefont {Landry},\ and\ \citenamefont {Li}}]{Hui_2021}%
  \BibitemOpen
  \bibfield  {author} {\bibinfo {author} {\bibfnamefont {L.}~\bibnamefont {Hui}}, \bibinfo {author} {\bibfnamefont {A.}~\bibnamefont {Joyce}}, \bibinfo {author} {\bibfnamefont {M.~J.}\ \bibnamefont {Landry}},\ and\ \bibinfo {author} {\bibfnamefont {X.}~\bibnamefont {Li}},\ }\href {https://doi.org/10.1088/1475-7516/2021/01/011} {\bibfield  {journal} {\bibinfo  {journal} {Journal of Cosmology and Astroparticle Physics}\ }\textbf {\bibinfo {volume} {2021}}\bibinfo  {number} { (01)},\ \bibinfo {pages} {011}}\BibitemShut {NoStop}%
\bibitem [{\citenamefont {Nikolaieva}\ \emph {et~al.}(2021)\citenamefont {Nikolaieva}, \citenamefont {Olashyn}, \citenamefont {Kuriatnikov}, \citenamefont {Vilchynskii},\ and\ \citenamefont {Yakimenko}}]{Nikolaieva}%
  \BibitemOpen
\bibfield  {number} {  }\bibfield  {author} {\bibinfo {author} {\bibfnamefont {Y.~O.}\ \bibnamefont {Nikolaieva}}, \bibinfo {author} {\bibfnamefont {A.~O.}\ \bibnamefont {Olashyn}}, \bibinfo {author} {\bibfnamefont {Y.~I.}\ \bibnamefont {Kuriatnikov}}, \bibinfo {author} {\bibfnamefont {S.~I.}\ \bibnamefont {Vilchynskii}},\ and\ \bibinfo {author} {\bibfnamefont {A.~I.}\ \bibnamefont {Yakimenko}},\ }\href {https://doi.org/10.1063/10.0005557} {\bibfield  {journal} {\bibinfo  {journal} {Low Temperature Physics}\ }\textbf {\bibinfo {volume} {47}},\ \bibinfo {pages} {684} (\bibinfo {year} {2021})}\BibitemShut {NoStop}%
\bibitem [{\citenamefont {Dmitriev}\ \emph {et~al.}(2021)\citenamefont {Dmitriev}, \citenamefont {Levkov}, \citenamefont {Panin}, \citenamefont {Pushnaya},\ and\ \citenamefont {Tkachev}}]{Dmitriev}%
  \BibitemOpen
  \bibfield  {author} {\bibinfo {author} {\bibfnamefont {A.}~\bibnamefont {Dmitriev}}, \bibinfo {author} {\bibfnamefont {D.}~\bibnamefont {Levkov}}, \bibinfo {author} {\bibfnamefont {A.}~\bibnamefont {Panin}}, \bibinfo {author} {\bibfnamefont {E.}~\bibnamefont {Pushnaya}},\ and\ \bibinfo {author} {\bibfnamefont {I.}~\bibnamefont {Tkachev}},\ }\bibfield  {journal} {\bibinfo  {journal} {Physical Review D}\ }\textbf {\bibinfo {volume} {104}},\ \href {https://doi.org/10.1103/physrevd.104.023504} {10.1103/physrevd.104.023504} (\bibinfo {year} {2021})\BibitemShut {NoStop}%
\bibitem [{\citenamefont {Zhang}\ \emph {et~al.}(2018)\citenamefont {Zhang}, \citenamefont {Chan}, \citenamefont {Harko}, \citenamefont {Liang},\ and\ \citenamefont {Leung}}]{zhang2018slowly}%
  \BibitemOpen
  \bibfield  {author} {\bibinfo {author} {\bibfnamefont {X.}~\bibnamefont {Zhang}}, \bibinfo {author} {\bibfnamefont {M.~H.}\ \bibnamefont {Chan}}, \bibinfo {author} {\bibfnamefont {T.}~\bibnamefont {Harko}}, \bibinfo {author} {\bibfnamefont {S.-D.}\ \bibnamefont {Liang}},\ and\ \bibinfo {author} {\bibfnamefont {C.~S.}\ \bibnamefont {Leung}},\ }\href@noop {} {\bibfield  {journal} {\bibinfo  {journal} {The European Physical Journal C}\ }\textbf {\bibinfo {volume} {78}},\ \bibinfo {pages} {1} (\bibinfo {year} {2018})}\BibitemShut {NoStop}%
\bibitem [{\citenamefont {Rindler-Daller}\ and\ \citenamefont {Shapiro}(2012)}]{rindler2012angular}%
  \BibitemOpen
  \bibfield  {author} {\bibinfo {author} {\bibfnamefont {T.}~\bibnamefont {Rindler-Daller}}\ and\ \bibinfo {author} {\bibfnamefont {P.~R.}\ \bibnamefont {Shapiro}},\ }\href@noop {} {\bibfield  {journal} {\bibinfo  {journal} {Monthly Notices of the Royal Astronomical Society}\ }\textbf {\bibinfo {volume} {422}},\ \bibinfo {pages} {135} (\bibinfo {year} {2012})}\BibitemShut {NoStop}%
\bibitem [{\citenamefont {Madarassy}\ and\ \citenamefont {Toth}(2013)}]{madarassy2013numerical}%
  \BibitemOpen
  \bibfield  {author} {\bibinfo {author} {\bibfnamefont {E.~J.}\ \bibnamefont {Madarassy}}\ and\ \bibinfo {author} {\bibfnamefont {V.~T.}\ \bibnamefont {Toth}},\ }\href@noop {} {\bibfield  {journal} {\bibinfo  {journal} {Computer Physics Communications}\ }\textbf {\bibinfo {volume} {184}},\ \bibinfo {pages} {1339} (\bibinfo {year} {2013})}\BibitemShut {NoStop}%
\bibitem [{\citenamefont {Silverman}\ and\ \citenamefont {Mallett}(2002)}]{silverman2002dark}%
  \BibitemOpen
  \bibfield  {author} {\bibinfo {author} {\bibfnamefont {M.~P.}\ \bibnamefont {Silverman}}\ and\ \bibinfo {author} {\bibfnamefont {R.~L.}\ \bibnamefont {Mallett}},\ }\href@noop {} {\bibfield  {journal} {\bibinfo  {journal} {General Relativity and Gravitation}\ }\textbf {\bibinfo {volume} {34}},\ \bibinfo {pages} {633} (\bibinfo {year} {2002})}\BibitemShut {NoStop}%
\bibitem [{\citenamefont {Zinner}(2011)}]{zinner2011vortex}%
  \BibitemOpen
  \bibfield  {author} {\bibinfo {author} {\bibfnamefont {N.~T.}\ \bibnamefont {Zinner}},\ }\href@noop {} {\bibfield  {journal} {\bibinfo  {journal} {Physics Research International}\ }\textbf {\bibinfo {volume} {2011}},\ \bibinfo {pages} {734543} (\bibinfo {year} {2011})}\BibitemShut {NoStop}%
\bibitem [{\citenamefont {Kain}\ and\ \citenamefont {Ling}(2010)}]{kain2010vortices}%
  \BibitemOpen
  \bibfield  {author} {\bibinfo {author} {\bibfnamefont {B.}~\bibnamefont {Kain}}\ and\ \bibinfo {author} {\bibfnamefont {H.~Y.}\ \bibnamefont {Ling}},\ }\href@noop {} {\bibfield  {journal} {\bibinfo  {journal} {Physical Review D}\ }\textbf {\bibinfo {volume} {82}},\ \bibinfo {pages} {064042} (\bibinfo {year} {2010})}\BibitemShut {NoStop}%
\bibitem [{\citenamefont {Rotha}\ and\ \citenamefont {Morgan}(2002)}]{rotha2002vortices}%
  \BibitemOpen
  \bibfield  {author} {\bibinfo {author} {\bibfnamefont {P.~Y.}\ \bibnamefont {Rotha}}\ and\ \bibinfo {author} {\bibfnamefont {M.~J.}\ \bibnamefont {Morgan}},\ }\href@noop {} {\bibfield  {journal} {\bibinfo  {journal} {Classical and Quantum Gravity}\ }\textbf {\bibinfo {volume} {19}},\ \bibinfo {pages} {L157} (\bibinfo {year} {2002})}\BibitemShut {NoStop}%
\bibitem [{\citenamefont {Toth}(2021)}]{Toth}%
  \BibitemOpen
  \bibfield  {author} {\bibinfo {author} {\bibfnamefont {V.~T.}\ \bibnamefont {Toth}},\ }\bibfield  {journal} {\bibinfo  {journal} {International Journal of Modern Physics D}\ }\textbf {\bibinfo {volume} {30}},\ \href {https://doi.org/10.1142/s0218271821501029} {10.1142/s0218271821501029} (\bibinfo {year} {2021})\BibitemShut {NoStop}%
\bibitem [{\citenamefont {{Mashhoon}}\ \emph {et~al.}(1984)\citenamefont {{Mashhoon}}, \citenamefont {{Hehl}},\ and\ \citenamefont {{Theiss}}}]{Hehl}%
  \BibitemOpen
  \bibfield  {author} {\bibinfo {author} {\bibfnamefont {B.}~\bibnamefont {{Mashhoon}}}, \bibinfo {author} {\bibfnamefont {F.~W.}\ \bibnamefont {{Hehl}}},\ and\ \bibinfo {author} {\bibfnamefont {D.~S.}\ \bibnamefont {{Theiss}}},\ }\href@noop {} {\bibfield  {journal} {\bibinfo  {journal} {General Relativity and Gravitation}\ }\textbf {\bibinfo {volume} {16}},\ \bibinfo {pages} {727} (\bibinfo {year} {1984})}\BibitemShut {NoStop}%
\bibitem [{\citenamefont {Medina}\ and\ \citenamefont {Gilmore}(2006)}]{Medina}%
  \BibitemOpen
  \bibfield  {author} {\bibinfo {author} {\bibfnamefont {J.}~\bibnamefont {Medina}}\ and\ \bibinfo {author} {\bibfnamefont {R.}~\bibnamefont {Gilmore}},\ }\href {https://books.google.de/books?id=R7BSNwAACAAJ} {\emph {\bibinfo {title} {Gravitoelectromagnetism (GEM): A Group Theoretical Approach}}}\ (\bibinfo  {publisher} {Drexel University},\ \bibinfo {year} {2006})\BibitemShut {NoStop}%
\bibitem [{\citenamefont {Mashhoon}(2003)}]{Mashhoon}%
  \BibitemOpen
  \bibfield  {author} {\bibinfo {author} {\bibfnamefont {B.}~\bibnamefont {Mashhoon}},\ }\href {https://doi.org/10.48550/ARXIV.GR-QC/0311030} {\bibinfo {title} {Gravitoelectromagnetism: A brief review}} (\bibinfo {year} {2003})\BibitemShut {NoStop}%
\bibitem [{\citenamefont {Wald}(1984)}]{Wald}%
  \BibitemOpen
  \bibfield  {author} {\bibinfo {author} {\bibfnamefont {R.~M.}\ \bibnamefont {Wald}},\ }\href {https://doi.org/10.7208/chicago/9780226870373.001.0001} {\emph {\bibinfo {title} {{General Relativity}}}}\ (\bibinfo  {publisher} {Chicago Univ. Pr.},\ \bibinfo {address} {Chicago, USA},\ \bibinfo {year} {1984})\BibitemShut {NoStop}%
\bibitem [{\citenamefont {Sarkar}\ \emph {et~al.}(2018)\citenamefont {Sarkar}, \citenamefont {Vaz},\ and\ \citenamefont {Wijewardhana}}]{sarkar2018gravitationally}%
  \BibitemOpen
  \bibfield  {author} {\bibinfo {author} {\bibfnamefont {S.}~\bibnamefont {Sarkar}}, \bibinfo {author} {\bibfnamefont {C.}~\bibnamefont {Vaz}},\ and\ \bibinfo {author} {\bibfnamefont {L.}~\bibnamefont {Wijewardhana}},\ }\href@noop {} {\bibfield  {journal} {\bibinfo  {journal} {Physical Review D}\ }\textbf {\bibinfo {volume} {97}},\ \bibinfo {pages} {103022} (\bibinfo {year} {2018})}\BibitemShut {NoStop}%
\bibitem [{\citenamefont {Cardoso}\ \emph {et~al.}(2022)\citenamefont {Cardoso}, \citenamefont {Ikeda}, \citenamefont {Vicente},\ and\ \citenamefont {Zilh{\~a}o}}]{cardoso2022parasitic}%
  \BibitemOpen
  \bibfield  {author} {\bibinfo {author} {\bibfnamefont {V.}~\bibnamefont {Cardoso}}, \bibinfo {author} {\bibfnamefont {T.}~\bibnamefont {Ikeda}}, \bibinfo {author} {\bibfnamefont {R.}~\bibnamefont {Vicente}},\ and\ \bibinfo {author} {\bibfnamefont {M.}~\bibnamefont {Zilh{\~a}o}},\ }\href@noop {} {\bibfield  {journal} {\bibinfo  {journal} {Physical Review D}\ }\textbf {\bibinfo {volume} {106}},\ \bibinfo {pages} {L121302} (\bibinfo {year} {2022})}\BibitemShut {NoStop}%
\bibitem [{\citenamefont {Duque}\ \emph {et~al.}(2023)\citenamefont {Duque}, \citenamefont {Macedo}, \citenamefont {Vicente},\ and\ \citenamefont {Cardoso}}]{duque2023axion}%
  \BibitemOpen
  \bibfield  {author} {\bibinfo {author} {\bibfnamefont {F.}~\bibnamefont {Duque}}, \bibinfo {author} {\bibfnamefont {C.~F.}\ \bibnamefont {Macedo}}, \bibinfo {author} {\bibfnamefont {R.}~\bibnamefont {Vicente}},\ and\ \bibinfo {author} {\bibfnamefont {V.}~\bibnamefont {Cardoso}},\ }\href@noop {} {\bibfield  {journal} {\bibinfo  {journal} {arXiv preprint arXiv:2312.06767}\ } (\bibinfo {year} {2023})}\BibitemShut {NoStop}%
\bibitem [{\citenamefont {Mitra}\ \emph {et~al.}(2023)\citenamefont {Mitra}, \citenamefont {Chakraborty}, \citenamefont {Vicente},\ and\ \citenamefont {Feng}}]{mitra2023probing}%
  \BibitemOpen
  \bibfield  {author} {\bibinfo {author} {\bibfnamefont {S.}~\bibnamefont {Mitra}}, \bibinfo {author} {\bibfnamefont {S.}~\bibnamefont {Chakraborty}}, \bibinfo {author} {\bibfnamefont {R.}~\bibnamefont {Vicente}},\ and\ \bibinfo {author} {\bibfnamefont {J.~C.}\ \bibnamefont {Feng}},\ }\href@noop {} {\bibfield  {journal} {\bibinfo  {journal} {arXiv preprint arXiv:2312.06783}\ } (\bibinfo {year} {2023})}\BibitemShut {NoStop}%
\bibitem [{\citenamefont {Doria}\ \emph {et~al.}(2007)\citenamefont {Doria}, \citenamefont {Romaguera},\ and\ \citenamefont {Peeters}}]{doria2007effect}%
  \BibitemOpen
  \bibfield  {author} {\bibinfo {author} {\bibfnamefont {M.~M.}\ \bibnamefont {Doria}}, \bibinfo {author} {\bibfnamefont {A.~R. d.~C.}\ \bibnamefont {Romaguera}},\ and\ \bibinfo {author} {\bibfnamefont {F.}~\bibnamefont {Peeters}},\ }\href@noop {} {\bibfield  {journal} {\bibinfo  {journal} {Physical Review B}\ }\textbf {\bibinfo {volume} {75}},\ \bibinfo {pages} {064505} (\bibinfo {year} {2007})}\BibitemShut {NoStop}%
\bibitem [{\citenamefont {Korshynska}\ \emph {et~al.}(2023)\citenamefont {Korshynska}, \citenamefont {Bidasyuk}, \citenamefont {Gorbar}, \citenamefont {Jia},\ and\ \citenamefont {Yakimenko}}]{korshynska2023dynamical}%
  \BibitemOpen
  \bibfield  {author} {\bibinfo {author} {\bibfnamefont {K.}~\bibnamefont {Korshynska}}, \bibinfo {author} {\bibfnamefont {Y.}~\bibnamefont {Bidasyuk}}, \bibinfo {author} {\bibfnamefont {E.}~\bibnamefont {Gorbar}}, \bibinfo {author} {\bibfnamefont {J.}~\bibnamefont {Jia}},\ and\ \bibinfo {author} {\bibfnamefont {A.}~\bibnamefont {Yakimenko}},\ }\href@noop {} {\bibfield  {journal} {\bibinfo  {journal} {The European Physical Journal C}\ }\textbf {\bibinfo {volume} {83}},\ \bibinfo {pages} {1} (\bibinfo {year} {2023})}\BibitemShut {NoStop}%
\bibitem [{\citenamefont {Inc.}()}]{Mathematica}%
  \BibitemOpen
  \bibfield  {author} {\bibinfo {author} {\bibfnamefont {W.~R.}\ \bibnamefont {Inc.}},\ }\href {https://www.wolfram.com/mathematica} {\bibinfo {title} {Mathematica, {V}ersion 14.0}},\ \bibinfo {note} {champaign, IL, 2024}\BibitemShut {NoStop}%
\bibitem [{\citenamefont {Jackson}\ \emph {et~al.}(1999)\citenamefont {Jackson}, \citenamefont {McCann},\ and\ \citenamefont {Adams}}]{jackson1999vortex}%
  \BibitemOpen
  \bibfield  {author} {\bibinfo {author} {\bibfnamefont {B.}~\bibnamefont {Jackson}}, \bibinfo {author} {\bibfnamefont {J.}~\bibnamefont {McCann}},\ and\ \bibinfo {author} {\bibfnamefont {C.}~\bibnamefont {Adams}},\ }\href@noop {} {\bibfield  {journal} {\bibinfo  {journal} {Physical Review A}\ }\textbf {\bibinfo {volume} {61}},\ \bibinfo {pages} {013604} (\bibinfo {year} {1999})}\BibitemShut {NoStop}%
\bibitem [{\citenamefont {McGee}\ and\ \citenamefont {Holland}(2001)}]{mcgee2001rotational}%
  \BibitemOpen
  \bibfield  {author} {\bibinfo {author} {\bibfnamefont {S.}~\bibnamefont {McGee}}\ and\ \bibinfo {author} {\bibfnamefont {M.}~\bibnamefont {Holland}},\ }\href@noop {} {\bibfield  {journal} {\bibinfo  {journal} {Physical Review A}\ }\textbf {\bibinfo {volume} {63}},\ \bibinfo {pages} {043608} (\bibinfo {year} {2001})}\BibitemShut {NoStop}%
\bibitem [{\citenamefont {Asakawa}\ and\ \citenamefont {Tsubota}(2024)}]{asakawa2024corotation}%
  \BibitemOpen
  \bibfield  {author} {\bibinfo {author} {\bibfnamefont {K.}~\bibnamefont {Asakawa}}\ and\ \bibinfo {author} {\bibfnamefont {M.}~\bibnamefont {Tsubota}},\ }\href@noop {} {\bibfield  {journal} {\bibinfo  {journal} {arXiv preprint arXiv:2409.07860}\ } (\bibinfo {year} {2024})}\BibitemShut {NoStop}%
\bibitem [{\citenamefont {Fetter}\ and\ \citenamefont {Svidzinsky}(2001)}]{fetter2001vortices}%
  \BibitemOpen
  \bibfield  {author} {\bibinfo {author} {\bibfnamefont {A.~L.}\ \bibnamefont {Fetter}}\ and\ \bibinfo {author} {\bibfnamefont {A.~A.}\ \bibnamefont {Svidzinsky}},\ }\href@noop {} {\bibfield  {journal} {\bibinfo  {journal} {Journal of Physics: condensed matter}\ }\textbf {\bibinfo {volume} {13}},\ \bibinfo {pages} {R135} (\bibinfo {year} {2001})}\BibitemShut {NoStop}%
\bibitem [{\citenamefont {Melia}\ \emph {et~al.}(2001)\citenamefont {Melia}, \citenamefont {Bromley}, \citenamefont {Liu},\ and\ \citenamefont {Walker}}]{Melia_2001}%
  \BibitemOpen
  \bibfield  {author} {\bibinfo {author} {\bibfnamefont {F.}~\bibnamefont {Melia}}, \bibinfo {author} {\bibfnamefont {B.~C.}\ \bibnamefont {Bromley}}, \bibinfo {author} {\bibfnamefont {S.}~\bibnamefont {Liu}},\ and\ \bibinfo {author} {\bibfnamefont {C.~K.}\ \bibnamefont {Walker}},\ }\href {https://doi.org/10.1086/320918} {\bibfield  {journal} {\bibinfo  {journal} {The Astrophysical Journal}\ }\textbf {\bibinfo {volume} {554}},\ \bibinfo {pages} {L37} (\bibinfo {year} {2001})}\BibitemShut {NoStop}%
\bibitem [{\citenamefont {Aschenbach}()}]{Aschenbach}%
  \BibitemOpen
  \bibfield  {author} {\bibinfo {author} {\bibfnamefont {B.}~\bibnamefont {Aschenbach}},\ }in\ \href {https://doi.org/10.1007/11403913_54} {\emph {\bibinfo {booktitle} {Growing Black Holes: Accretion in a Cosmological Context}}}\ (\bibinfo  {publisher} {Springer-Verlag})\ pp.\ \bibinfo {pages} {302--303}\BibitemShut {NoStop}%
\bibitem [{\citenamefont {Reynolds}(2021)}]{Reynolds_2021}%
  \BibitemOpen
  \bibfield  {author} {\bibinfo {author} {\bibfnamefont {C.~S.}\ \bibnamefont {Reynolds}},\ }\href {https://doi.org/10.1146/annurev-astro-112420-035022} {\bibfield  {journal} {\bibinfo  {journal} {Annual Review of Astronomy and Astrophysics}\ }\textbf {\bibinfo {volume} {59}},\ \bibinfo {pages} {117} (\bibinfo {year} {2021})}\BibitemShut {NoStop}%
\bibitem [{\citenamefont {Groszek}\ \emph {et~al.}(2018)\citenamefont {Groszek}, \citenamefont {Paganin}, \citenamefont {Helmerson},\ and\ \citenamefont {Simula}}]{groszek2018motion}%
  \BibitemOpen
  \bibfield  {author} {\bibinfo {author} {\bibfnamefont {A.~J.}\ \bibnamefont {Groszek}}, \bibinfo {author} {\bibfnamefont {D.~M.}\ \bibnamefont {Paganin}}, \bibinfo {author} {\bibfnamefont {K.}~\bibnamefont {Helmerson}},\ and\ \bibinfo {author} {\bibfnamefont {T.~P.}\ \bibnamefont {Simula}},\ }\href@noop {} {\bibfield  {journal} {\bibinfo  {journal} {Physical Review A}\ }\textbf {\bibinfo {volume} {97}},\ \bibinfo {pages} {023617} (\bibinfo {year} {2018})}\BibitemShut {NoStop}%
\end{thebibliography}%

\end{document}